\documentclass{egpubl}
\usepackage{pg2022}

\SpecialIssueSubmission    

\usepackage[T1]{fontenc}
\usepackage{dfadobe}  
\usepackage{wrapfig}

\usepackage[T1]{fontenc}
\usepackage{dfadobe}  
\usepackage{amsmath,amssymb,amsfonts}%
\usepackage{mathrsfs}%
\usepackage{algorithm}%
\usepackage{algorithmicx}%
\usepackage{algpseudocode}%
\usepackage{array}

\biberVersion
\BibtexOrBiblatex
\usepackage[backend=biber,bibstyle=EG,citestyle=alphabetic,backref=true]{biblatex} 
\electronicVersion
\PrintedOrElectronic
\ifpdf \usepackage[pdftex]{graphicx} \pdfcompresslevel=9
\else \usepackage[dvips]{graphicx} \fi

\usepackage{egweblnk} 
\usepackage{multirow}
\usepackage{wrapfig,graphicx,lipsum}

\title[NSTO: Neural Synthesizing Topology Optimization for Modulated Structure Generation]%
      {NSTO: Neural Synthesizing Topology Optimization\\for Modulated Structure Generation}

\author[S. Zhong \& P. Punpongsanon \& D. Iwai \& K. Sato]
{\parbox{\textwidth}{\centering Shengze Zhong\orcid{0000-0003-3295-0564}, Parinya Punpongsanon\orcid{0000-0003-2720-7768}, Daisuke Iwai\orcid{0000-0002-3493-5635}, Kosuke Sato\orcid{0000-0003-1429-9990}}
        \\
{\parbox{\textwidth}{\centering Graduate School of Engineering Science, Osaka University, Toyonaka, Japan
       }
}
}

\begin{document}

\teaser{
 \includegraphics[width=\linewidth]{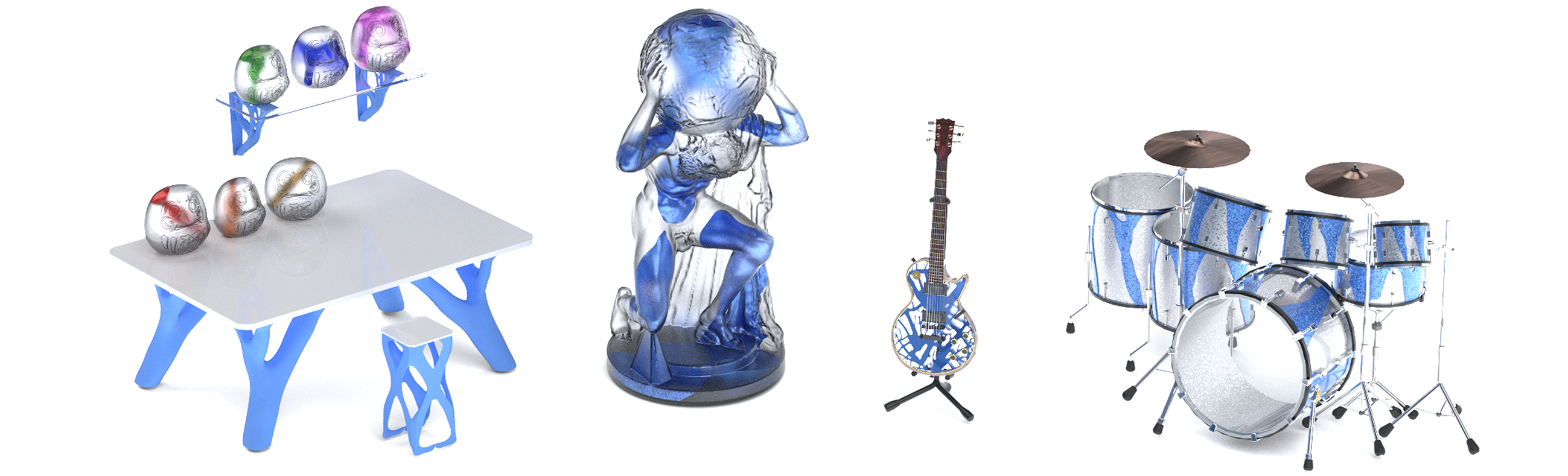}
 \centering
  \caption{Structures with maximized stiffness and limited material consumption are optimized with low computation cost, all shown in color. The proposed NSTO also enables generating interpolated solutions with an instant network feedforward, like the seven drum shells.}
\label{fig:teaser}
}

\maketitle
\begin{abstract}
   Nature evolves structures like honeycombs at optimized performance with limited material. These efficient structures can be artificially created with the collaboration of structural topology optimization and additive manufacturing. However, the extensive computation cost of topology optimization causes low mesh resolution, long solving time, and rough boundaries that fail to match the requirements for meeting the growing personal fabrication demands and printing capability. Therefore, we propose the neural synthesizing topology optimization that leverages a self-supervised coordinate-based network to optimize structures with significantly shorter computation time, where the network encodes the structural material layout as an implicit function of coordinates. Continuous solution space is further generated from optimization tasks under varying boundary conditions or constraints for users' instant inference of novel solutions. We demonstrate the system's efficacy for a broad usage scenario through numerical experiments and 3D printing.

\ccsdesc[500]{Applied computing~Computer-aided design}
\ccsdesc[500]{Computing methodologies~Shape analysis}
\ccsdesc[300]{Computing methodologies~Computer graphics}

\printccsdesc   
\end{abstract}  
\section{Introduction}
The evolution of nature has created various elaborate structures, such as lotus leaves and honeycombs, whose objective performance is maximized at an acceptable cost. Being enlightened, humans have created magnificent structures, from wagon wheels to Sagrada Familia. The study of optimal structural design finally evolves to topology optimization with growing algorithms and computation power.

Structural topology optimization is a mathematical method that maximizes the structure performance under constraints by reasonably distributing the material layout. 
\setlength{\columnsep}{3pt}
\begin{wrapfigure}{r}{3.2cm}
\label{wrap-fig:1}
\vspace{-4.5mm}
\hspace{0.6mm}
\includegraphics[width=\linewidth]{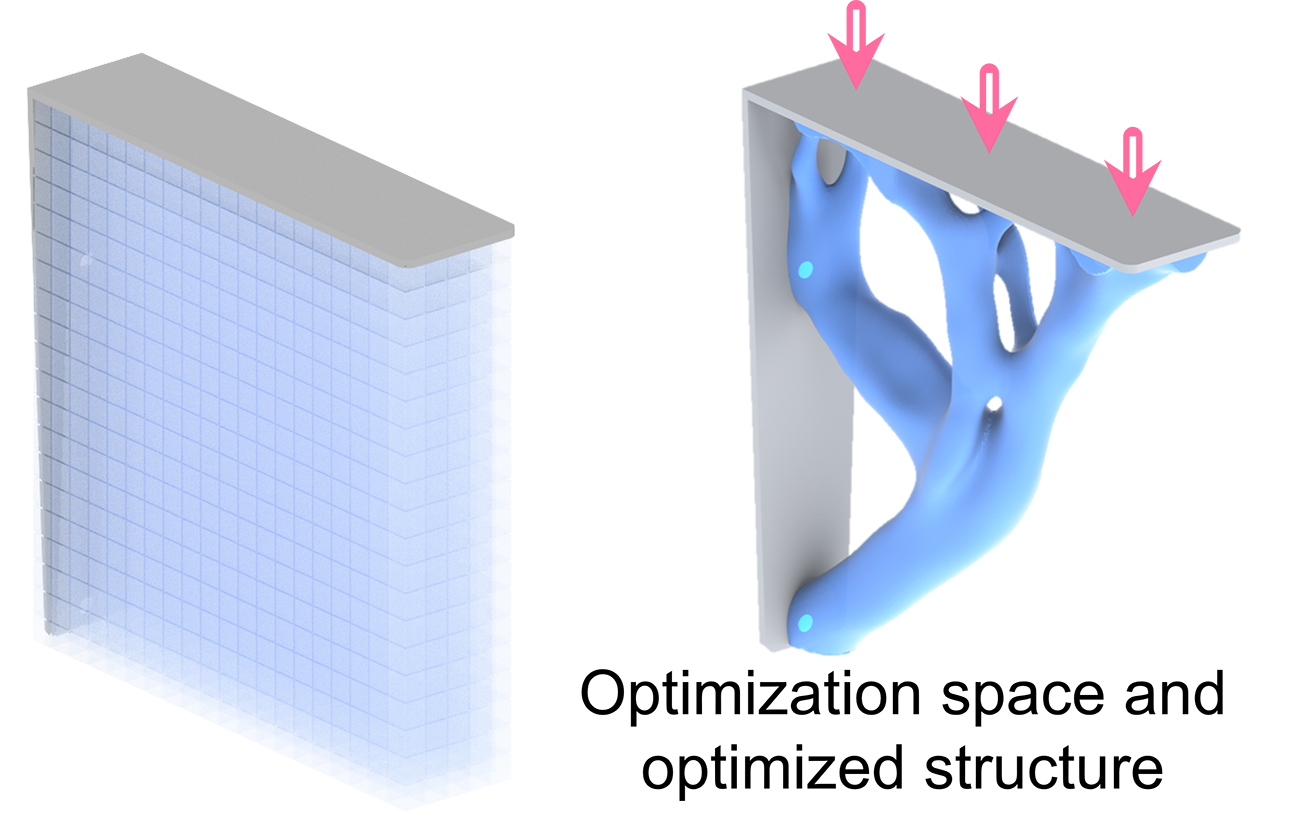}
\vspace{-10mm}
\end{wrapfigure}
Specifically, the optimization space is discretized into elements and iteratively filled or deprived of materials to approach design objects. While topology optimization has broad industrial applications in mechanics and architecture \cite{bendsoe2003topology}, its exorbitant computational cost hinders its popularity for personal fabrication among ordinary users, especially those without high-performance hardware or cloud servers. To achieve fine structures with smooth boundaries, computation at high-performance workstations extend from hours to days \cite{aage2017giga, liu2018narrow}. This situation worsens when frequent design-parameter adjustments are required.

We proposed the neural synthesizing topology optimization (NSTO) to address these two problems for ordinary users, with the following physical-informed dual networks: \textbf{\textit{a:}} a coordinate-based network for implicit-neural-represented structural topology optimization for computing fine and smooth structures with competitive performance and a much shorter time; \textbf{\textit{b:}} a self-supervised auto-decoder network for further generating a series of structures under continuous boundary conditions for user selection. 
The dual networks are named \textit{oscillator network} and \textit{modulator network} respectively, since NSTO shares a similar workflow with an analog music synthesizer for generating, modulating, and filtering a structure (waveform). Such network design has been proven to have higher reconstruction quality than a single coordinate-based network for image and shape representation \cite{mehta2021modulated}. Figure \ref{fig:teaser} shows some of the structures optimized by the proposed system.

\textit{Oscillator network} is a coordinate-based network that represents and optimizes a single structure at an arbitrary resolution. It inputs coordinates and outputs the material spatial layout so that concatenated network outputs can implicitly form a structure. The physical loss function is obtained by performing finite element analysis on the network outputted structure, and its gradient is leveraged for network updating. At the inference time, a super-resolution (i.e., voxel-wise interpolation) is performed to resample the structure representation to an arbitrary higher resolution.

\textit{Modulator network} is an auto-decoder network that enhances the overall network expressiveness by layer-wisely modulating the oscillator network's weight. It is only activated to optimize multiple structures under varying boundary conditions. Within, multiple optimization subtasks are simultaneously carried out, and we expect the network to learn interval solutions among subtasks. For example, the user may input subtasks for $30\%$ and $70\%$ structure volume constraints and generate a novel solution of a $50\%$ volume constraint. Each subtask is assigned a latent code, and these latent codes are randomly initialized, inputted, and updated with the network weight. During training, the latent codes are gradually clustered to separate positions in the latent space, labeling the corresponding boundary condition. Finally, the user can interpolate the latent codes to infer novel structures in a corresponding solution space.

We assert the rationality and necessity of NSTO from three aspects. Generally, structural topology optimization is a non-convex problem where its flat solution space leads to a series of near-optimal solutions \cite{sigmund1998numerical}. Such property provides adequate space for new algorithm development, focusing either on performance \cite{wu2015system}, manufacture \cite{bi2020topology} or aesthetics \cite{hu2019texture}. Technically, there is a disequilibrium between the slight structural performance enhancement and computation cost of high-resolution topology optimization, which indicates that multi-resolution structure representation is a potential solution \cite{zhu2017two}. Practically, users are also concerned about affordable computation cost and good structure appearance besides high-level performance.

We compare NSTO with both physical and data-driven methods and demonstrate its competitive optimization results and much shorter computation time for high-resolution structures. As a physics-informed system, it performs self-supervision with no dataset generation cost and no data-driven artifacts. Besides comparing NSTO with benchmark methods, a series of applications are presented, including the usage and numerical results. Finally, we 3D-printed the optimized structures to further verify the feasibility.

The main contributions are as follows:

\begin{itemize}
\item To improve topology optimization efficiency, the oscillator network optimizes and super-resolves structures in an implicit neural form, shortening computation from hours/minutes to seconds.
\item To improve solution space generation performance, the modulator network enhances the overall expressiveness, enabling novel solution generation through feedforwarding latent codes.
\end{itemize}

\section{Related works}\label{sec2}

In this section, we summarize the studies related to our method on the development and computation efficiency of topology optimization and the neural field techniques that the proposed topology optimization framework builds upon.

\subsection{Topology optimization}\label{subsec2.1}

Topology optimization aims to enhance structural performance by redistributing the spatial layout of the material. Its traditional applications include, but are not limited to stiffness enhancement \cite{buhl2000stiffness}, natural frequency reduction \cite{pedersen2000maximization}, and heat conduction \cite{gersborg2006topology}, and they have been broadening to personal fabrications of interactive toys, furniture, musical instruments, and so on.

Topology optimization has multiple mainstream algorithms. The explicit methods, such as the solid isotropic material with penalization (SIMP) \cite{bendsoe1989optimal} and bidirectional evolutionary structural optimization (BESO) \cite{huang2007convergent} discretize the optimization space into structured grids. These methods feature simple algorithms and fast convergence but result in manufacturing difficulties because of fuzzy boundaries. The implicit method based on a level set \cite{wang2003level} parameterizes the structure as a contour to achieve smooth boundaries, but it needs to reset the level set equation to ensure continuous updating, which causes a computation burden. Moreover, topology optimization is unstable, as it includes checkerboard patterns, mesh dependencies, and local minima. On the contrary, however, these features widen the algorithm design space by enabling near-optimal options of structures with similar performance, such as using texture-driven topology optimization with personalized designs \cite{hu2019texture}, and bone-like infill optimization \cite{wu2017infill}.

\begin{figure*}[ht]
    \centering
        \includegraphics[width=\linewidth]{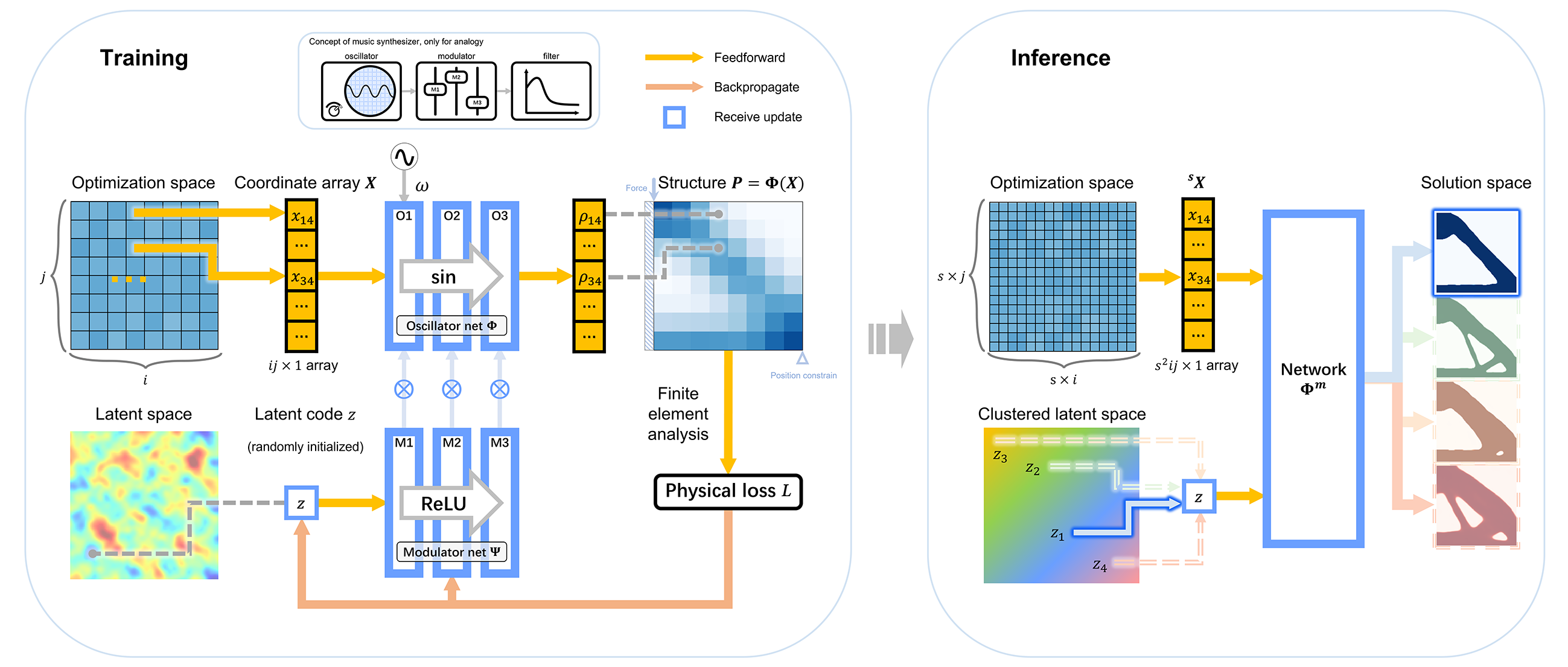}
    \caption{NSTO system diagram. For single structure optimization, the oscillator network imports coordinate array $\textit{\textbf{X}}$ from optimization space and exports structure $\textit{\textbf{P}}$, finite element analysis is performed to get the loss $L$. For solution space generation, the modulator network activates. Multiple subtasks are simultaneously optimized, and their corresponding latent codes $z$ are randomly initialized, input, and optimized with network parameters. At inference, the user may super-resolve the solutions by taking a denser grid and generating novel solutions through feedforwarding interpolated latent code. The upper icons are for analogy.}
    \label{fig:systemoverview}
\end{figure*}

Large-scale topology optimization at high resolution has always been a challenging problem \cite{mukherjee2021accelerating}. \cite{aage2017giga} realized giga-voxel resolution plane wing topology optimization with the parallel operation of supercomputers in a few days. \cite{wu2015system} developed a high-performance GPU solver based on the algebraic multigrid method to improve the solving efficiency while ensuring convergence, thereby realizing million-voxel resolution optimization in a few minutes with a personal computer. \cite{liu2018narrow} proposed a narrow-band topology optimization that focuses the computation on the generated thin structure area instead of equally solving the entire optimization space. They successfully optimized a billion-voxel bird beak structure using a workstation in 113.19 hours.

\subsection{Neural field}\label{subsec2.2}

Neural computing provides powerful tools for improving computation efficiency. One of the main-stream trends is training end-to-end data-driven networks for non-iterative topology optimization \cite{zhang2019deep, wang2020machine}. However, the generalization and interpretability of data-driven networks are still doubted, and their optimization quality is highly subject to the training data.

The recent development of the neural field \cite{xie2021neural} has ignited new opportunities in fields such as visual computing, signal processing, and physical simulation. A neural field, or coordinate-based neural network, is a neural implicit re-parameterization of signals. The network inputs a discrete sampling position and outputs the signal sampling value (e.g., image RGB value and audio waveform). It can date back to the compositional pattern producing networks (CPPN) \cite{stanley2007compositional} in 2007; and after a series of fruitful research and development since 2019, it can now represent high-fidelity signals \cite{tancik2020fourier, sitzmann2020implicit, martel2021acorn}. In the successive studies of \cite{sitzmann2020implicit, tancik2020fourier}, the Fourier domain neural field demonstrated improved high-frequency expression, which also enabled frequency adjustment of the network output \cite{doosti2021topology}.

As for neural represented topology optimization, \cite{hoyer2019neural} proactively proposed that the neural re-parameterization of topologies improved the optimization quality. \cite{chandrasekhar2021tounn, zhang2021tonr} further verified various network models with classical tasks. \cite{zehnder2021ntopo} proposed the converse density space objective, which applies the step-wise predicted structure to guide the network fitting.

Recent studies on implicit-neural-represented structural topology optimization were generally at an early stage and usually did not address practical issues such as optimization quality and computation cost. First, on the optimization quality, previous networks struggled to optimize high-frequency structure details \cite{deng2020topology, chandrasekhar2021tounn}, which are especially important for structures under distributed external force. Second, the optimization convergence was unsatisfactory, as hundreds of iterations were required for a classic benchmark task \cite{zhang2021tonr} (Messerschmitt-Bölkow-Blohm beam), while conventional explicit methods cost dozens of iterations \cite{huang2007convergent, ferrari2020new}. Although several studies have demonstrated exciting progress, such as with the mesh-independent FEA solver (still tradeoff with the computation time) \cite{zehnder2021ntopo}, the overall effects are doubted due to the deficiency in the verification and comparison of the optimization-performance under complex tasks.

In conclusion, topology optimization provides a potent tool for high-quality structure design. However, its enormous computation hinders its popularity among ordinary users. To achieve the overall balance between the computation time and the structure quality, we contribute to a neural topology optimization framework for efficiently generating near-optimal structures with an appreciable reduction of time, and also enabling instant generation of novel solutions by performing multi-object optimization and latent space interpolation. Comprehensive comparison and various tasks were carried out to support the above statement.

\section{Neural synthesizing topology optimization}\label{sec3}

NSTO consists of three parts:
\begin{itemize}
\item Sec. \ref{subsec3.1}: A physical solver that adopts finite element analysis (FEA) to compute the structure performance.
\item Sec. \ref{subsec3.2}: An oscillator network that represents the structure in an implicit neural form for super-resolution.
\item Sec. \ref{subsec3.3}: A modulator network that enhances the network expressiveness for multi-structure generation.
\end{itemize}

The system overview based on these three sections is briefly shown in Equation \ref{eq0} and Figure \ref{fig:systemoverview}. First, the oscillator network $\Phi$ inputs an $n\times3$ coordinates array $X$ to generate an initial structure $P$. Second, the physical loss function $L$ of the structure $P$ is computed from the FEA solver. Third, the gradient of loss $L$ is backpropagated as $\textrm{BP}(L)$ to update the network $\Phi$. After the iterations of the above steps, the user may infer a super-resolved structure $^{s}\textbf{\textit{P}}$ by inputting the coordinates array of an $s$-times denser grid $^{s}\textbf{\textit{X}}$, similar to the image super-resolution methods \cite{chen2021learning}.
\begin{gather}
\label{eq0}
    \left.\
    \begin{array}{ll}
    Training:
        \left\{
            \begin{array}{lr}
                \textbf{\textit{P}} = \Phi(\textbf{\textit{X}}) \\
                L = \textrm{FEA}(\textbf{\textit{P}}) \\
                \Phi \Leftarrow \textrm{BP}(L)  \\
            \end{array}
        \right. \\
        Inference: \; \; \; {^{s}\textbf{\textit{P}}} = \Phi(^{s}\textbf{\textit{X}}) \\
    \end{array} \
    \right.
\end{gather}

The modulator network is activated to optimize multiple structures under varying boundary conditions or constraints (e.g., volume constraints) to interpolate solutions. The dual networks (i.e., oscillator and modulator) $\Phi^m$ input a latent code $\textbf{\textit{z}}$ with a coordinate array $\textbf{\textit{X}}$ and generate a corresponding structure as $\textbf{\textit{P}}_i = {\Phi^m}(\textbf{\textit{X}}, \textbf{\textit{z}}_i)$, where $i$ is the index of the subtasks. Procedures in Equation \ref{eq0} are repeated to compute the loss and update the network and latent code.

\subsection{Physical solver}\label{subsec3.1}

This section establishes the physical loss function $L = \textrm{FEA}(\textbf{\textit{P}})$ and its gradient, and then introduce parameters required for FEA.

\textit{Structure} $\textbf{\textit{P}}$ is represented by an $i \times j$ matrix $\textbf{\textit{P}}_{ij}=\rho_{ij}^{\tau}$, according to the Solid Isotropic Material with Penalization (SIMP) method \cite{bendsoe1989optimal}. Each of its element density $\rho \in [0,1]$ tells the material existence at that position. To comprehend this better, the reader may imagine a binary image $\textbf{\textit{P}}$ whose binary pixel values $\rho$ describe the shape. The penalty factor $\tau$ is to enhance the convergence of the material density to binary values for clearly judging if the material should be allocated to that position \cite{sigmund1998numerical}, since an intermediate density such as $0.5$ is practically meaningless in manufacturing.

\textit{Compliance} $C$ is selected as the optimization object in this study. Structure compliance minimization is a popular topic for designing lightweight and strong structures with minimal deformation under the same volume and load. As shown in Equation \ref{eq2}, $C$ is the sum of all the finite elements' compliance weighted by $\textbf{\textit{P}}$ and formed by the structure deformation $\textbf{\textit{U}}$ and the element stiffness matrix $\textbf{\textit{K}}_e$. A volume constraint is imposed, where $V(\textbf{\textit{P}})=\sum \textbf{\textit{P}}$ is the structure volume, $V_0$ is the $100\%$-filled volume of the optimization space, and $\delta$ is the volume fraction.
\begin{gather}
\label{eq2}
    \left.\
    \begin{array}{ll}
    \underset{\textbf{\textit{P}}}{\mathrm{argmin}}\, C(\textbf{\textit{P}}) = \sum\textbf{\textit{P}} {\textbf{\textit{U}}^T}\textbf{\textit{K}}_e\textbf{\textit{U}} \\
        s.t. \left.\
    \begin{array}{ll}
    V(\textbf{\textit{P}})/V_0 \le \delta
    \end{array}
    \right.
    \end{array}
    \right.
\end{gather}

\textit{Loss function} $L$ is constructed using the Augmented Lagrangian method \cite{fortin2000augmented}, which transfers the volume-constrained compliance minimization problem into unconstrained form, as shown in Equation \ref{eq:lagrangian}. $\lambda$ is the Lagrangian multiplier that updates as $\lambda_{k+1} = \lambda_{k} + 2\sigma_{k} (\frac{V(\textbf{\textit{P}})}{V_0} - \delta)^2$, and $\sigma_{k}$ is the penalty factor that exponentially grows with the iteration $k$, whose base number is empirically set as $1.1$. A mean-squared-error volume constraint $(\frac{V(\textbf{\textit{P}})}{V_0} - \delta)^2$ is adopted to avoid convergence oscillations. Thus, the Lagrange multiplier and the penalty term are quadratic and fourth-power.
\begin{equation}
\label{eq:lagrangian}
    \underset{\textbf{\textit{P}}}{\mathrm{argmin}}\, L(\textbf{\textit{P}}) = C(\textbf{\textit{P}}) + \lambda (\frac{V(\textbf{\textit{P}})}{V_0} - \delta)^2 + 
    \sigma (\frac{V(\textbf{\textit{P}})}{V_0} - \delta)^4
\end{equation}

\textit{Loss gradient} $\frac{\partial L}{\partial \rho}$ with respect to each element's density is computed as shown in Equation \ref{eq:gradient}, where the volume gradient $\frac{dV}{d\rho}=1$. In Sec. \ref{subsec3.2}, this loss gradient will be used to perform network backpropagation based on the chain rule.
\begin{equation}
\label{eq:gradient}
    \frac{\partial L}{\partial \rho} = -\tau\rho^{\tau-1}\textbf{\textit{U}}^T\textbf{\textit{K}}_e\textbf{\textit{U}} + 2\lambda (\frac{V(\textbf{\textit{P}})}{V_0} - \delta) + 
    4\sigma (\frac{V(\textbf{\textit{P}})}{V_0} - \delta)^3
\end{equation}

With the loss function and its gradient established, we now introduce the necessary physical parameters for loss computation, following the standard FEA process.

First, we assemble the structure stiffness matrix $\textbf{\textit{K}}$. The element stiffness matrix $\textbf{\textit{K}}_e=\int_{v_e}\textbf{\textit{B}}^T\textbf{\textit{c}}\textbf{\textit{B}}d{v_e}$ is computed from the strain matrix $\textbf{\textit{B}}$, the material constitutive model $\textbf{\textit{c}}$ and the element volume $v_e$. In NSTO, we respectively apply the rectangular and hexahedron elements to 2D and 3D cases, each element having its specific $\textbf{\textit{B}}$, $\textbf{\textit{c}}$, and $v_e$. $\textbf{\textit{K}}$ is thus assembled from $\textbf{\textit{K}}_e$ by adding the $\textbf{\textit{K}}_e$ of each element to the corresponding position with the same degree-of-freedom index in the blank $\textbf{\textit{K}}$ matrix \cite{rao2017finite}.

Second, we obtain the structure deformation $\textbf{\textit{U}}$ (i.e., element-nodal-wise displacement) by solving the large-scale linear equation, Equation \ref{eq:linearsolve}, where $\textbf{\textit{F}}$ is the external force. This is the most computationally expensive step in compliance-minimization topology optimization, as it usually takes up to $75\%$ to $95\%$ computation time in both the state-of-the-art and our methods. The GPU-based algebraic multigrid method (AMGX) \cite{naumov2015amgx} is applied for acceleration, as it is $1$ to $2$ orders of magnitude faster than the analytical solver in our experiment setup (Sec. \ref{subsec4.1}). Specifically, the preconditioner is set as a multi-layer v-hierarchy with the Jacobi smoother. The convergence tolerance and the max iteration are set at $1\textrm{e}-8$ and $100$, respectively, for stable convergence in various tasks. Although the AMGX setup can be further optimized, we emphasize that NSTO focuses on the efficiency improvement brought by the implicit-neural-representation framework rather than by the linear equation solver, and there is no contradiction between the combined implementation of NSTO and advanced numerical solvers.
\begin{equation}
\label{eq:linearsolve}
  \textbf{\textit{KU}} = \textbf{\textit{F}}
\end{equation}

In conclusion, the physical loss function $L$ and its gradient for training the networks are established and computed with an accelerated algebraic multigrid solver.

\begin{figure*}
    \centering
    \includegraphics[width=\linewidth]{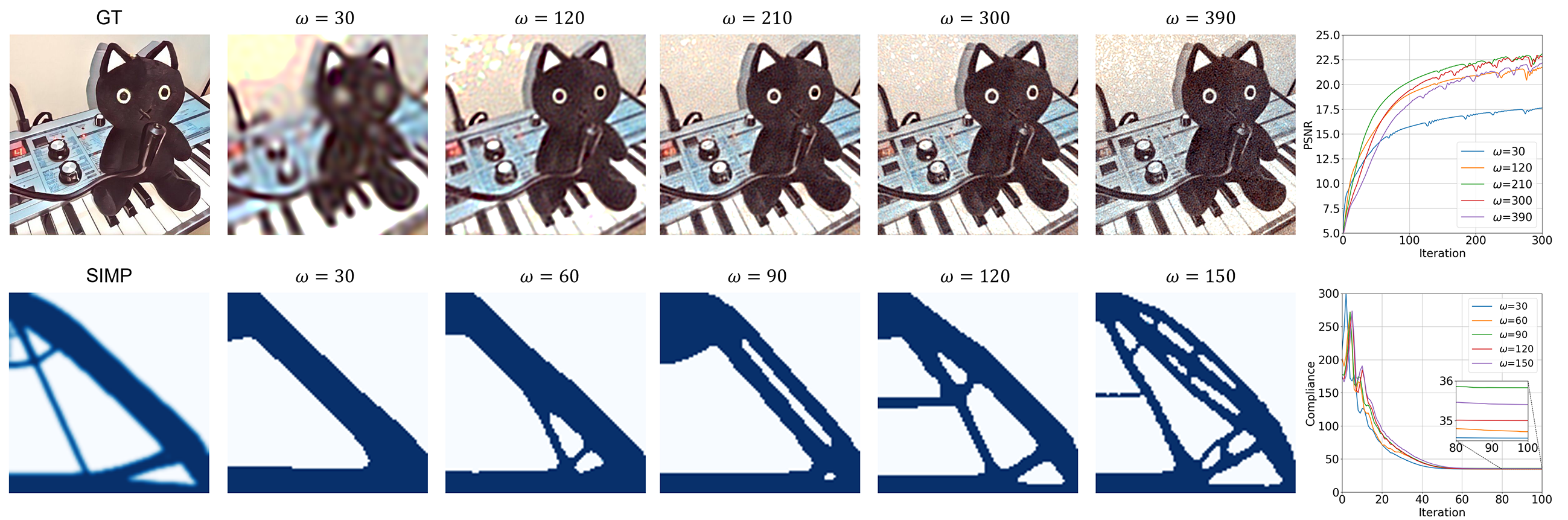}
    \caption{The effect of the network frequency tuning mechanism on image fitting and topology optimization. High-frequency image feature increases with the ascending hyperparameter $\omega$, as with the structures. The solution qualities are measured with PSNR and compliance.}
    \label{fig:frequencymodulation}
\end{figure*}

\subsection{Oscillator network}\label{subsec3.2}

This section focuses on the oscillator network that reparameterizes the structure material layout with coordinates, as $\textbf{\textit{P}} = \Phi(\textbf{\textit{X}})$. The structure super-resolution and frequency tuning mechanisms are subsequently introduced.

\textit{The oscillator network} applies a SIREN network \cite{sitzmann2020implicit}, namely, a sinusoidal activated multilayer perceptron \cite{gardner1998artificial}. It has been proven capable of complex signal representation \cite{mildenhall2020nerf, zhang2021editable} and has significantly stronger high-frequency expressiveness than ReLU-activated networks. Equation \ref{eq:layer} shows the network's layer-wise output $\phi_i(\textbf{\textit{x}})$, where $\textbf{\textit{w}}$, $\textbf{\textit{b}}$, and $i$ are the neuron's weight, bias, and the layer index, respectively. The final layer of the network standardizes its output to the $[0,1]$ interval for the material density $\rho$, as $\textrm{Atan}(\textbf{\textit{x}})=\pi^{-1}\arctan{\alpha \textbf{\textit{x}}}+0.5$. $\alpha$ is a hyperparameter that adjusts the sensitive interval of the network outputs and is empirically set at $\alpha = 0.1$ for smoother convergence.
\begin{equation}
\label{eq:layer}
    \phi_i(\textbf{\textit{x}}) = \sin({\textbf{\textit{w}}_i}{\boldsymbol{\phi}_{i-1}\textbf{\textit{x}}}+\textbf{\textit{b}}_i), \ \textbf{\textit{w}}_1=\omega\boldsymbol{\phi}_{0}
\end{equation}

Therefore, the oscillator network inputs coordinate $x$ and outputs material density $\rho$, creating an in-between analog relationship, as shown in Equation \ref{eq:oscnetIO}. In practice, a full-batch training $\textbf{\textit{P}}=\Phi(\textbf{\textit{X}})$ is applied for faster gradient descent. All the sampled coordinates $x$ are assembled as a $(i\times j \times k, 3)$ array in 3D cases, where $i, j, k$ are the axial resolutions of the optimization space. The network and its output structure are updated once per epoch. 
\begin{equation}
\label{eq:oscnetIO}
    {\rho}=\Phi(\textbf{\textit{x}})=\textrm{Atan}[{\textbf{\textit{w}}_i}{(\boldsymbol{\phi}_{i-1}\circ\cdots\circ\boldsymbol{\phi}_0)(\textbf{\textit{x}})+\textbf{\textit{b}}_i}]
\end{equation}

The loss function gradient of the net parameters is obtained through the chain rule as $\partial L/ \partial \textbf{\textit{w}} = (\partial L/\partial \rho)\times(\partial\rho/\partial\textbf{\textit{w}})$. $\partial L/\partial\textbf{\textit{w}}$ is computed with Autograd \cite{paszke2017automatic}. $\partial L/\partial \rho$ is manually computed to avoid the conspicuous gradient computing costs due to the iterative solving of $\textbf{\textit{KU}}=\textbf{\textit{F}}$ using the algebraic multigrid method.

Among the feasible optimizers including Adam \cite{kingma2014adam}, L-BFGS-B \cite{zhu1997algorithm}, and resilient backpropagation (Rprop) \cite{riedmiller1993direct}, we selected Rprop due to its faster convergence in most tasks, despite its lower convergence stability compared to that of Adam. To balance the convergence speed and quality, we also set the penalty factor $\tau \in [1.5, 3]$, as shown in Equation \ref{eq:gradient}.

At the inference stage, the user may discretize the optimization space into an $s$-times denser grid $^{s}\textbf{\textit{X}}$ to obtain the $s$-times super-resolved structure $^{s}\textbf{\textit{P}}$ as Figure \ref{fig:systemoverview}. One may imagine discretizing an image with a smaller pixel size so that its boundary is smoothed.
\begin{equation}
\label{eq6}
    ^{s}\textbf{\textit{P}} = \Phi(^{s}\textbf{\textit{X}})
\end{equation}

The Fourier-featured network further enables the structure frequency tuning mechanism, which positively relates the structure frequency to the network hyperparameter $\omega$ initialized in the first layer, as $\textbf{\textit{w}}_1=\omega\boldsymbol{\phi}_{0}$. This mechanism enables the generation of high-frequency structural details that are typically obtained at high-resolution computation, thereby improving the robustness against local structural failures. Users can also reduce the structural frequency for higher manufacturability. In practice, the maximal $\omega$ is limited to avoid generating invalid structures during the super-resolution, since an excessive frequency will cause signal aliasing and worsen structural optimality. For an analogy, we demonstrate this mechanism with the image fitting task shown in Figure \ref{fig:frequencymodulation}, where the network inputs pixel coordinates, outputs RGB values, and is trained to fit a ground truth image with the mean-squared error. We observed that the fitted image moved from underfitting to overfitting when $\omega$ was increased.

In conclusion, the oscillator network implicitly represents a structure material layout by relating the element density $\rho$ to its coordinate $x$, thereby bringing the functions of structure super-resolution and frequency tuning. There are also several differences from explicit topology optimization methods. First, the initial element density is randomly distributed around $\rho=0.5$, which probably converges to more local minima (i.e., optimized structures) than the uniformly initialized density in conventional methods. Second, the density filter in conventional methods \cite{bourdin2001filters} is spared since the implicit neural representation has a similar effect, which is discussed from the perspective of the neural tangent kernel \cite{dupuis2021dnn}.

\subsection{Modulator network}\label{subsec3.3}

This section focuses on the modulator network, which is used to enhance the performance of multi-structure optimization under varying boundary conditions so that novel structures can be generated in boundary-condition intervals.  

Structure generation is the second research object that we hope to address besides optimization efficiency. In practical scenarios, users often adjust design parameters repetitively for desired structures, and each adjustment requires a complete optimization. For this reason, the network is expected to optimize multiple solutions under different boundary conditions labeled with the corresponding latent code $z$, so that new solutions can be instantly generated by interpolating the latent space (i.e., the latent codes interval).

To this end, structures with different boundary conditions are simultaneously optimized for generating interpolated solutions. Different boundary conditions or constraints are reflected in the loss function $L$, as shown in Equation \ref{eq:lagrangian}. Here, we introduce two tasks. First, a varying volume constraint is realized by assigning different volume fractions $\delta$ in the loss function $L$. Second, a varying external force is realized by adjusting the force $\textbf{\textit{F}}$ in Equation \ref{eq:linearsolve} for steering the compliance $C$ in the loss function $L$. For example, two structures with latent codes $z_1, z_2$ and volume fractions $\delta_1=0.3, \delta_2=0.7$ are assembled as a ‘dataset’ and alternately optimized. Users are expected to obtain a structure with a volume of $\frac{V}{V_0}=0.5$ through feedfowarding an interpolated latent code $\frac{z_1+z_2}{2}$. In brief, the multi-structure generation task is to minimize a varying loss function that caters to different boundary conditions. 

\begin{wrapfigure}{r}{3.3cm}
\label{wrap-fig:2}
\vspace{-4mm}
\hspace{0.5mm}
\includegraphics[width=\linewidth]{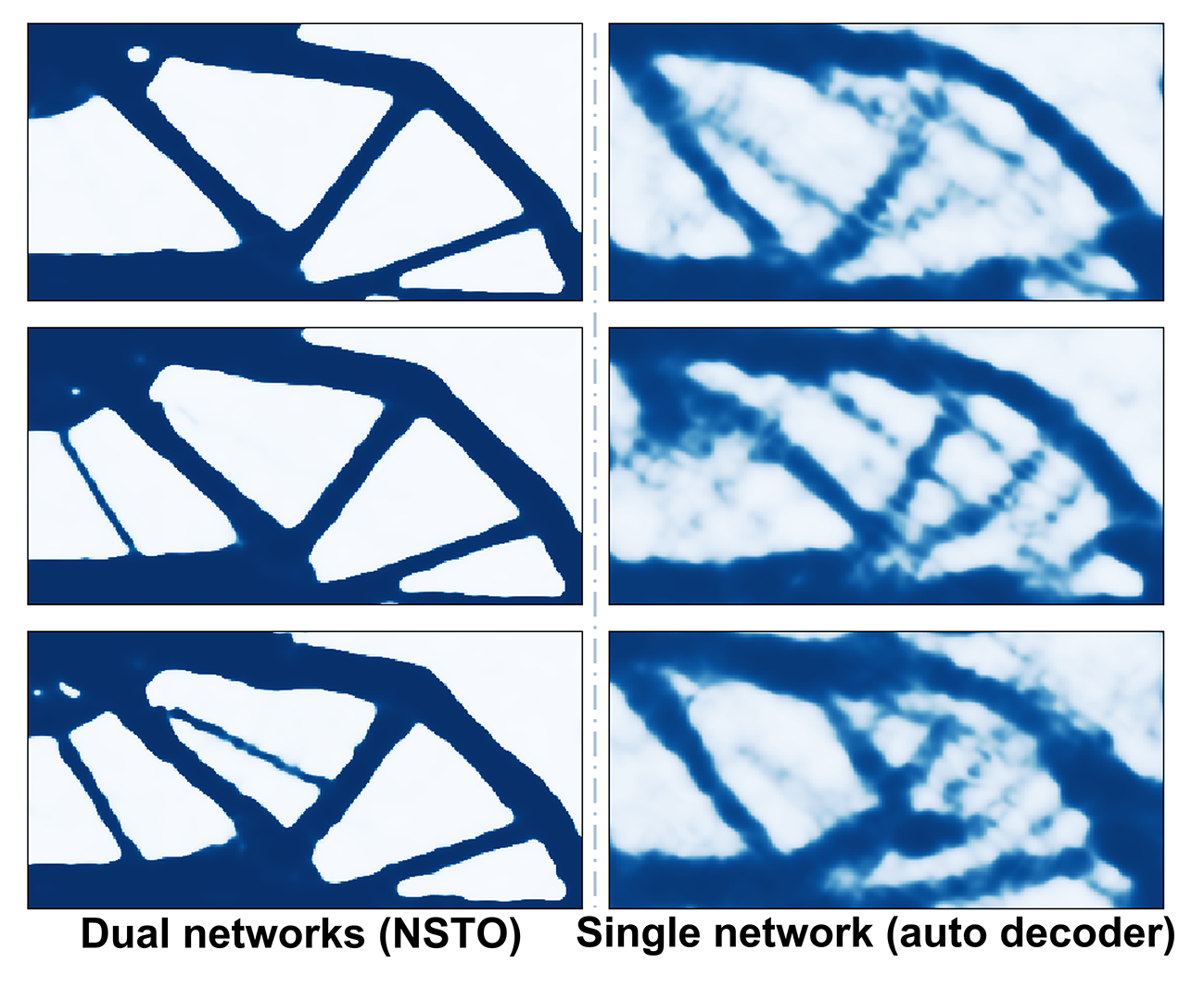}
\vspace{-8mm}
\end{wrapfigure}
We attempt to use a single oscillator network as an autodecoder \cite{park2019deepsdf} for the task. Autodecoder is a multilayer perceptron that inputs concatenated coordinates and latent code $(x, z)$, and outputs the density $\rho$, as $\rho = \Phi(x,z)$. During the training, the randomly initialized latent code is updated together with the network parameters so that the encoding procedure in a typical variational autoencoder \cite{kingma2013auto} is spared. At the inference stage, users may feedforward interpolated latent codes for novel structure generation. The autodecoder is simple in design and efficient in training. However, we found that a single autodecoder struggles with the multi-structure generation task, mainly due to the lack of network expressiveness.

\textit{The modulator network} is thus applied to promote the network performance. It is a ReLU-activated multilayer perceptron $\Psi$ with a size identical to the oscillator. It applies the same strategy as that with an autodecoder, as the latent code is initialized with a normal distribution and updated with the network parameters like a 'layer0', thus sparing the encoder. The latent code is used to modulate the oscillator network to enhance network expressiveness \cite{mehta2021modulated}, rather than generating structures itself. Equation \ref{eq:modulator} shows the output of the layer $\boldsymbol{\psi}_{i}(z)$ and the network $\Psi(z)$, where $z$ is the latent code and $\textbf{\textit{w}}^{'}$ and $\textbf{\textit{b}}^{'}$ are the layer's weight and bias.  
\begin{gather}
\label{eq:modulator}
    \left.\
    \begin{array}{ll}
    \psi_i(\textbf{\textit{z}}) = \textrm{ReLU}(\textbf{\textit{w}}_i^{'}\boldsymbol{\psi}_{i-1}\textbf{\textit{z}}+\textbf{\textit{b}}_i^{'}) \\
    \Psi(\textbf{\textit{z}})=\textbf{\textit{w}}_i^{'}(\boldsymbol{\psi}_{i-1}\circ\cdots\circ\boldsymbol{\psi}_0)(\textbf{\textit{z}})+\textbf{\textit{b}}_i^{'} \\
    \end{array}
    \right.
\end{gather}

\begin{algorithm}[t]
\caption{Neural synthesizing topology optimization}\label{algo}
\begin{algorithmic}[1]
\State \textbf{Input:} \:$\textbf{\textit{X}}, \textbf{\textit{bc}}$ \quad \# coordinate array and boundary condition
\State \textbf{Output:} \:$^{s}\textbf{\textit{P}}$ \,\,\quad \# $s$-times super-resolved structure

\\
\State \# Physical solver
\algrenewcommand{\Return}{\State\algorithmicreturn~}
\Function{FEA}{\textbf{\textit{X, bc}}}
    \State $\textbf{\textit{F}}, \delta, = \textbf{\textit{bc}}$
    \State $B, c, v_e = \textrm{ElemParas}()$  \quad\# element-dependent
    \State $\textbf{\textit{K}}_e=\int_{v_e}\textbf{\textit{B}}^T\textbf{\textit{c}}\textbf{\textit{B}}d{v_e}$
    \State $\textbf{\textit{K}}$ = assemble($\textbf{\textit{K}}_e$)
    \State $U = \textrm{AMG}(\textbf{\textit{K}}, \textbf{\textit{F}})$
    \State $C = \sum\textbf{\textit{P}} {\textbf{\textit{U}}^T}\textbf{\textit{K}}_e\textbf{\textit{U}}$
    \State $L = C +\lambda(\frac{V(\textbf{\textit{P}})}{V_0} - \delta)^2 + \sigma (\frac{V(\textbf{\textit{P}})}{V_0} - \delta)^4$
    \Return $L, \frac{\partial L}{\partial \rho}$
\EndFunction

\\
\State \# Single-structure optimization 
\Procedure{TrainOscNet}{\textbf{\textit{X, bc}}}
    \State $\textrm{OscNet.initialize}()$ \quad \# OscNet $\Phi$
    \State $\textrm{optimizer.initialize}()$
    \For{$epoch$ in $\textrm{range}(e_{max})$}
        \State $\textbf{\textit{P}}=\Phi(\textbf{\textit{X}})$  
        \State $L=\textrm{FEA}(\textbf{\textit{P}}, \textbf{\textit{bc}})$
        \State $\Phi \Leftarrow \textrm{BP}(L)$ \quad
        \State $\textrm{optimizer.step}()$
    \EndFor
    \Return $^{s}\textbf{\textit{P}} = \Phi(^{s}\textbf{\textit{X}})$ 
\EndProcedure

\\
\State \# Multi-structure optimization, the main algorithm
\Procedure{TrainModOscNet}{\textbf{\textit{X, bc}}}
    \State $\textrm{OscNet.initialize}()$ \quad \# OscNet $\Phi$
    \State $\textrm{ModNet.initialize}()$ \:\: \# Modulated OscNet $\Phi^m$
    \State $\textrm{optimizer.initialize}()$
    \For{$epoch$ in $\textrm{range}(e_{max})$}
        \For{$i$ in $\textrm{enumerate}(subtasks)$}
            \State $\textbf{\textit{P}}_i = \Phi^{m}(\textbf{\textit{X}}, \textbf{\textit{z}}_i)$
            \State $L=\textrm{FEA}(\textbf{\textit{P}}_i, \textbf{\textit{bc}}_i)$
            \State $\Phi^m \Leftarrow \textrm{BP}(L)$
        \State $\textrm{optimizer.step}()$
        \EndFor
    \EndFor
    \Return $^{s}\textbf{\textit{P}} = \Phi^{m}(^{s}\textbf{\textit{X}}, \textbf{\textit{z}})$
\EndProcedure
\end{algorithmic}
\end{algorithm}
\bigskip

It enhances the expressiveness of the oscillator network through element-wise multiplication of each of its layers' outputs by the corresponding outputs of the oscillator, which are marked with the symbol \textcircled{$\times$} in Figure \ref{fig:systemoverview}. For comparison, a single autodecoder that inputs concatenated latent code and coordinates $\textbf{\textit{P}} = \Phi(\textbf{\textit{X}}, \textbf{\textit{z}})$ only causes a phase shift to the layer's output, as $\phi_i(\textbf{\textit{x}}) = \sin({\textbf{\textit{w}}_{i,x}}{\boldsymbol{\phi}_{i-1}\textbf{\textit{x}}}+{\textbf{\textit{w}}_{i,z}}{\boldsymbol{\phi}_{i-1}\textbf{\textit{x}}}+\textbf{\textit{b}}_i)$, so its expressiveness is weaker.

The dual networks $\Phi^{m}$ take the coordinate $x$ and the latent code $z$ respectively, and outputs the material density $\rho$ of the $z$-labeled structure, as Equation \ref{eq:modNet}. Through the feature maps of the two networks in training, we observed that both networks learn the optimal structure in cooperation. At inference, users can perform a simple network feedforward $\textbf{\textit{P}}= \Phi^{m}(\textbf{\textit{X}}, \textbf{\textit{z}})$ to infer a structure or perform an interpolation of latent codes for new designs.
\begin{gather}
\label{eq:modNet}
    \left.\
    \begin{array}{ll}
    \phi^{m}_i(\textbf{\textit{x}}, \textbf{\textit{z}}) = \psi_i(\textbf{\textit{z}})\sin({\textbf{\textit{w}}_i}{\boldsymbol{\phi}^{m}_{i-1}\textbf{\textit{x}}}+\textbf{\textit{b}}_i) \\
    \rho=\Phi^{m}(\textbf{\textit{x}}, \textbf{\textit{z}})=\textrm{Atan}[{\psi_i(\textbf{\textit{z}})}{(\boldsymbol{\phi}_{i-1}\circ\cdots\circ\boldsymbol{\phi}_0)(\textbf{\textit{x}})+\textbf{\textit{b}}_i}] \\
    \end{array}
    \right.
\end{gather}

In conclusion, the modulator network is introduced to enhance the network's expressiveness. Therefore, the dual networks (oscillator and modulator) are capable of structure generation tasks under varying boundary conditions, enabling users to generate intermediate solutions by interpolating and feedforwarding the latent codes of optimized subtasks. The pseudo code of NSTO is shown in Algorithm \ref{algo}, where function FEA summarizes the physical solver (Sec. \ref{subsec3.1}), and the two procedures below summarize the training steps of the oscillator (Sec. \ref{subsec3.2}) and the dual networks (Sec. \ref{subsec3.3}).

\section{Experiments}\label{sec4}

NSTO is experimentally introduced and verified from three aspects:
\begin{itemize}
\item Sec. \ref{subsec4.1}: The experimental setup is briefly introduced.
\item Sec. \ref{subsec4.2}: The numerical performance of NSTO, including the convergence rate, optimization quality (compliance), and time, are demonstrated. The benchmark algorithms are compared.
\item Sec. \ref{subsec4.3}: Various applications are shown with instructions. Structures are 3D-printed for manufacturability verification.
\end{itemize}

\subsection{Experiment setup}\label{subsec4.1}

This section introduces the experimental environment, the network hyperparameters, and the training method. 

NSTO experiments run on a desktop PC (Intel Core i9-10900KF, Nvidia RTX2080S 8GB, Ubuntu 16.04, Python 3.6), where GPU determines the network scale and FEA resolution.

The network hyperparameters were set as follows. The network width was commonly set at $512$ for balanced VRAM consumption and performance, and the learning rate was fixed at $1\mathrm{e}{-4}$. The hyperparameter $\omega$ for structure frequency tuning was commonly set at $60$ and could be adjusted according to the resolution and optimization object. We empirically set $\omega$ less than twice the maximum axial resolution to avoid the aliasing caused by an exorbitant frequency. The latent code dimension was set at 1D and 2D for the intuitive inference of the latent space and can be slid up for challenging tasks. 

Full-batch training was performed by assembling all the discretized coordinates of the optimization space as an array and feeding it to the network as a whole so that the network outputs would be the complete element density values of a structure. The feedforwarding, FEA, backpropagation, and network updating are performed once per epoch. Generally, optimization converges after tens of epochs, depending on the loading condition.

\subsection{Performance}\label{subsec4.2}

\begin{figure}[ht]
    \centering
    \includegraphics[width=\linewidth]{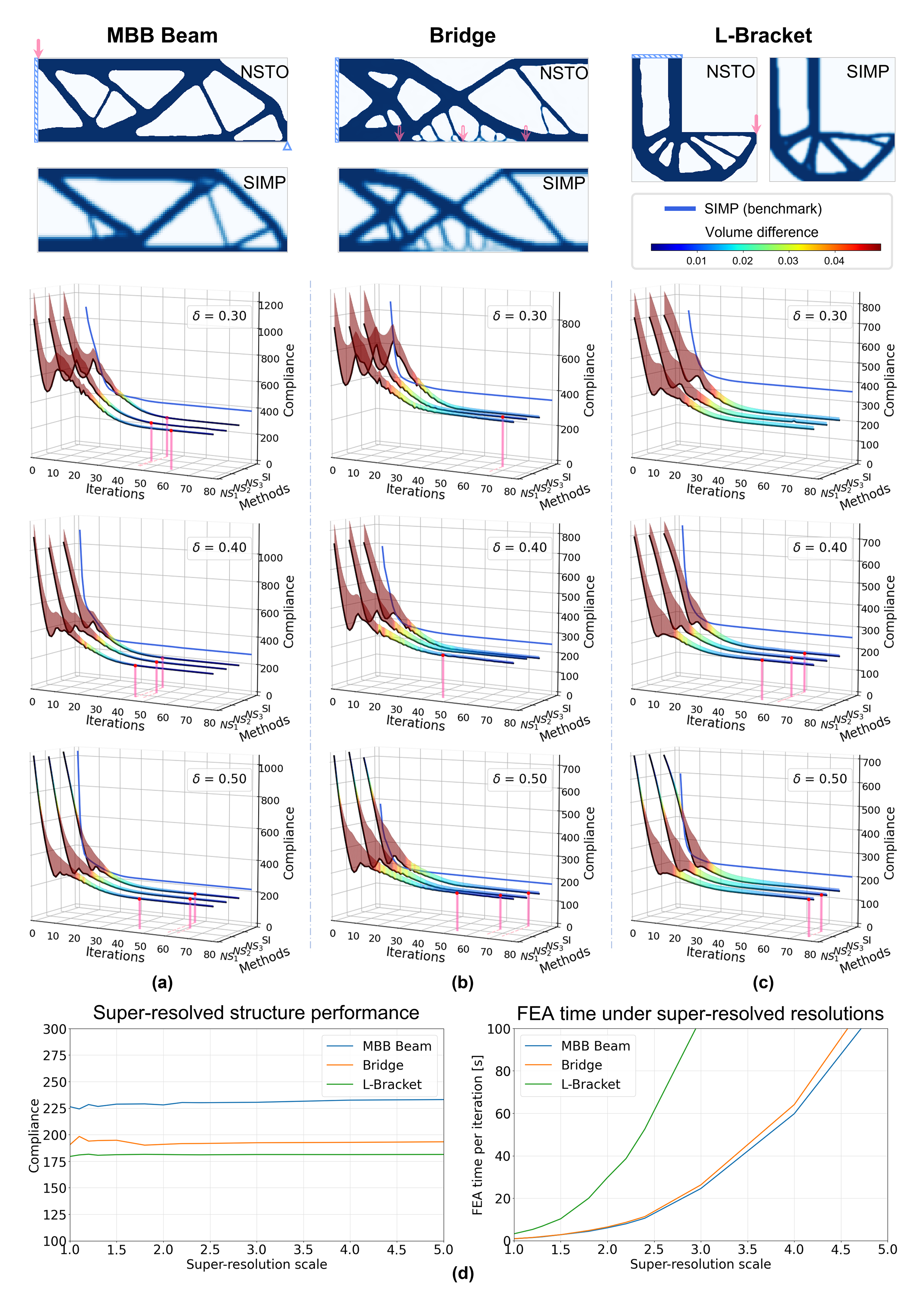}
    \caption{Convergence and performance of NSTO, compared with SIMP in (a) MBB beam, (b) Bridge, and (c) L-Bracket optimization, where optimized structures are shown above the convergence plots. Red arrows indicate external force, and blue marks indicate fixed position constraints. (d) Performance of several super-resolved NSTO results and the ordinary FEA iteration time under the same resolutions.}
    \label{fig:convergence}
\end{figure}  

\newcolumntype{P}[1]{>{\centering\arraybackslash}p{#1}}
\begin{table}[ht]
\caption{NSTO and SIMP benchmark solutions in 80 iterations. $C$ is the compliance, $V$ and $\delta$ are the optimized and constrained volume fraction.}\label{tbl1}
\resizebox{\linewidth}{!}{

\begin{tabular}{@{}lllllll@{}}
\hline
&\multicolumn{2}{c}{MBB Beam}
&\multicolumn{2}{c}{Bridge}
&\multicolumn{2}{c}{L-Bracket}\\\hline
Method&C$\downarrow$&V$\downarrow$ ($\delta$)&C$\downarrow$&V$\downarrow$ ($\delta$)&C$\downarrow$&V$\downarrow$ ($\delta$)\\\hline  

NSTO $_{(\omega=60)}$&292.88&0.31 (0.30)&287.30&0.31 (0.30)&\textbf{223.05}&0.32 (0.30)\\
NSTO $_{(\omega=120)}$&\textbf{287.18}&0.31 (0.30)&279.65&0.31 (0.30)&223.39&0.31 (0.30)\\
NSTO $_{(\omega=180)}$&297.33&0.30 (0.30)&\textbf{267.09}&0.31 (0.30)&225.82&0.31 (0.30)\\
SIMP&373.89&0.30 (0.30)&343.37&0.30 (0.30)&352.10&0.30 (0.30)\\\hline

NSTO $_{(\omega=60)}$&224.44&0.41 (0.40)&207.95&0.41 (0.40)&180.08&0.42 (0.40)\\
NSTO $_{(\omega=120)}$&226.36&0.41 (0.40)&208.18&0.41 (0.40)&\textbf{179.46}&0.42 (0.40)\\
NSTO $_{(\omega=180)}$&\textbf{221.80}&0.41 (0.40)&\textbf{190.48}&0.41 (0.40)&181.35&0.42 (0.40)\\
SIMP&271.00&0.40 (0.40)&241.05&0.40 (0.40)&249.39&0.40 (0.40)\\\hline

NSTO $_{(\omega=60)}$&188.46&0.51 (0.50)&164.23&0.51 (0.50)&155.69&0.51 (0.50)\\
NSTO $_{(\omega=120)}$&186.69&0.51 (0.50)&150.46&0.51 (0.50)&155.58&0.51 (0.50)\\
NSTO $_{(\omega=180)}$&\textbf{184.79}&0.51 (0.50)&\textbf{149.47}&0.51 (0.50)&\textbf{155.06}&0.52 (0.50)\\
SIMP&213.80&0.50 (0.50)&174.21&0.50 (0.50)&198.46&0.50 (0.50)\\\hline

\end{tabular}}
\end{table}

This section introduces the numerical performance of NSTO from two aspects: \textbf{\textit{a:}} its convergence and structural performance compared to those of the benchmark method; \textbf{\textit{b:}} computation time for tasks with complex structures and boundary conditions.

\begin{figure*}
    \centering
    \includegraphics[width=\linewidth]{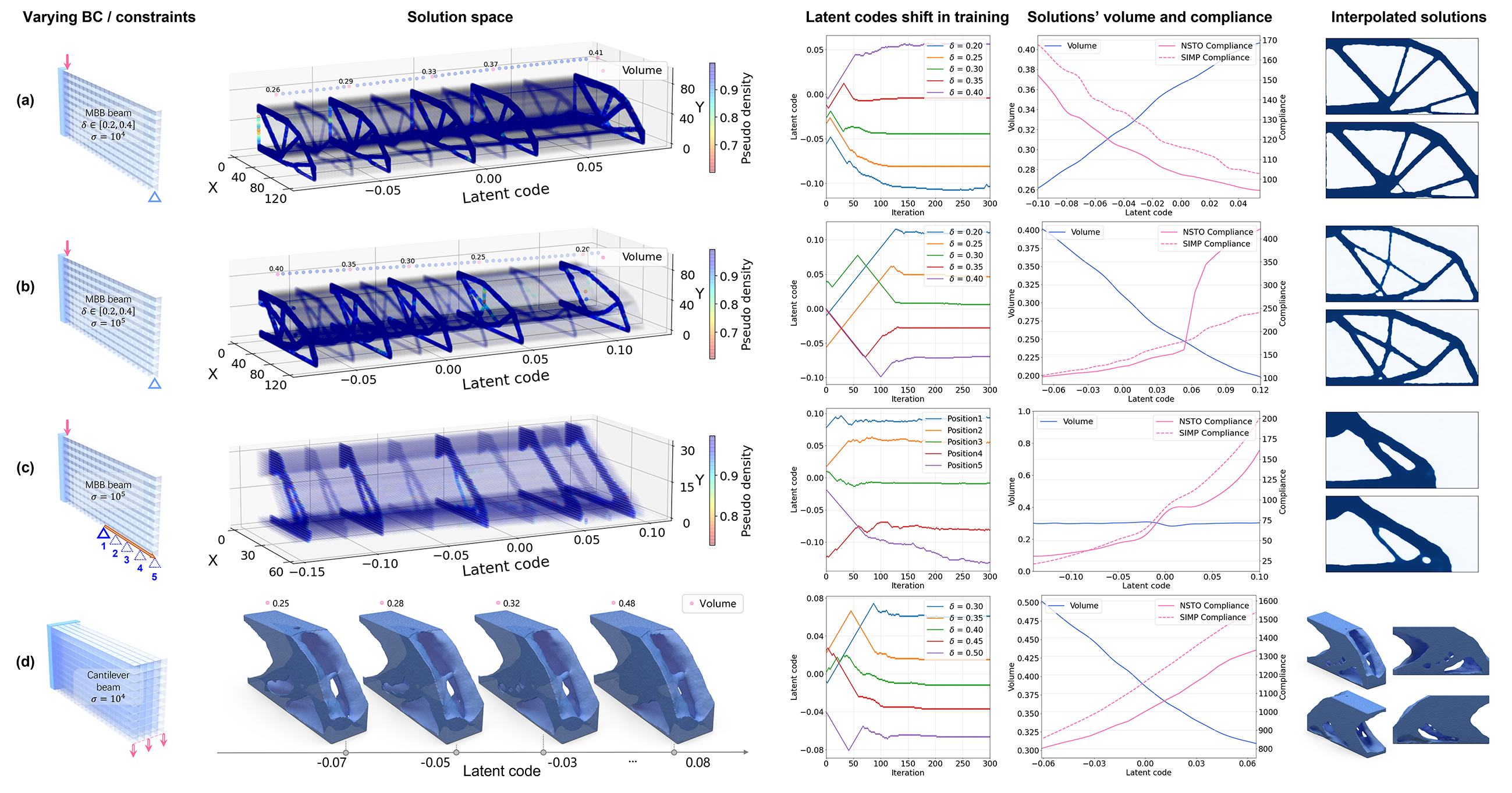}
    \caption{Generated solution space of (a, b) varying volume-constrained MBB beams with different penalty factor $\sigma$; (c) varying supporting-position MBB beams; and (d) varying volume-constrained cantilever beams. Original subtasks are shown in solid lines, and interpolated solutions are shown in semi-transparent. The upper dots indicate the volume, and the color in the solution space indicates the pseudo density. The boundary conditions or constraints are shown in 3D for better visualization.}
    \label{fig:solutionspace}
\end{figure*}

First, the convergence and the optimized structural performance (compliance) of NSTO and the benchmark SIMP are compared \cite{ferrari2020new}. The Solid Isotropic Material with Penalization (SIMP) is a structural topology optimization algorithm that is widely applied in academies and industries for its simple implementation, fast convergence, and capability to optimize non-linear geometries and materials. It is also generally referenced as an algorithm benchmark. For fairness, we used the same algebraic-multigrid-method FEA solver \cite{bell2022pyamg} for its stable solutions, computed in the same environment, and set up optimal configurations for SIMP by tuning the density filter radius. Note that the different initialization of NSTO and SIMP is only a derivative result of the algorithms' formulation of constrained optimization rather than being a decisive factor of structural optimality. Specifically, the initialization strategy of SIMP comes from the tight volume constraint imposed by its optimality criteria method, which is not necessary for NSTO.

\textit{Convergence} of NSTO was tested in three benchmark structural topology optimization scenarios \cite{valdez2017topology}: the optimization of the Messerschmitt-Bölkow-Blohm (MBB) beam, bridge, and L-Bracket. Their boundary conditions covered the ranges of the concentrated and distributed external force and the irregular optimization space, as shown in the top part of Figure \ref{fig:convergence}. Specifically, the MBB beam was fixed at two bottom ends and received a top-central concentrated force. The bridge structure was also fixed at two ends but subjected to uniformly distributed force on the lower surface. The L-Bracket was optimized in an L-sized optimization space, where it was fixed at the top and received a bending force on the right. Conventionally, the first two tasks leveraged symmetric constraints for computation efficiency, so half of these structures were optimized. During the experiments, the MBB beam and bridge were optimized at a $120\times 40$ resolution, and the L-Bracket was optimized at a $100 \times 100$ resolution. The detail boundary condition setting followed the benchmark \cite{valdez2017topology}.

The convergence results under different volume fractions $\delta$ and frequency hyperparameters $\omega$ are enumerated in the middle of Figure \ref{fig:convergence}. The black lines and the rainbow-colored strips indicate the compliance and the volume loss at the current iteration, respectively. The blue lines indicate the compliance of the benchmark SIMP method. The red dots and the pink lines indicate the convergence iteration of NSTO, whose criteria are set as a $0.3\%$ compliance difference from the previous iteration and a $1\%$ volume difference from the fraction. The results show that NSTO under different $\omega$ values converged towards optimal solutions at a stable pace.

\textit{Performance} of the structures were comprehensively compared. As shown in Figure \ref{fig:convergence} and Table \ref{tbl1}, NSTO had lower compliance than SIMP in all the benchmark cases. The clear structure boundaries of the NSTO solutions also led to minor efforts in mesh post-processing. As for the volume difference with the volume fraction, NSTO had a $1\%$ error mainly due to the loose volume constraints from the augmented Lagrangian method. Besides the $1\times$ resolution structure comparison, we super-resolved the solutions to observe their shape and performance change. The structures were super-resolved $10$ times, but no artifacts appear, as shown in the top of Figure \ref{fig:convergence}. Then, we performed FEA on the super-resolved structure and observed that the compliance remained stable in the tested range ($1-5\times$ axial super-resolution), which verified the feasibility of implicit-neural-represented structure optimization. Compared with the negligible time for structure super-resolution (i.e., a simple network feedforward), performing FEA at the same resolution was much more time-consuming, as shown in the bottom right part of Figure \ref{fig:convergence}. The above results verify the first objective of NSTO, that is, to improve the computation efficiency of structural topology optimization for ordinary users. We re-emphasize that NSTO contributes to the overall structure optimization framework rather than to the numerical FEA solver on which it is built.

Second, the dual networks were examined in three solution space generation tasks: the varying volume-constrained MBB beam task MBB$_{v}$, the varying supporting-position MBB beam task MBB$_{s}$, and the varying volume-constrained 3D cantilever beam task CB$_{v}$, as shown in Figure \ref{fig:solutionspace}. The latent codes of all the tasks were set at 1D to directly view the solution space $P={\Phi^{m}}(\textit{\textbf{X}}, z)$ that corresponded to the latent codes $z: \mathbb{R}^1$.

The MBB$_{v}$ task adopted varying volume constraints to generate MBB beams with intermediate volume as Figure \ref{fig:solutionspace} (a, b). To this end, the network simultaneously trained five optimization subtasks with equally distributed volume constraints as $\delta \in [0.20, 0.25, 0.30, 0.35, 0.40]$. Note that the loss function $L$ was subject to varying constraints $\delta$, which are necessary for other 1D manipulations such as varying external force. During the training, the latent codes of each subtask gradually mapped to the 1D latent-space positions in the order of volume constraint $\delta$ and presented an approximately linear relationship with $\delta$. We uniformly sampled along the $z$ axis at the inference stage and computed the volume and compliance of the corresponding output structure. Most interpolation solutions perform better than SIMP, consistent with the observations in Figure \ref{fig:convergence}. We also note that when an excessively large penalty factor $\sigma$ (Equation \ref{eq:lagrangian}) was imposed on the volume constraint, there was a trade-off of optimality, which was manifested as ascending compliance in Figure \ref{fig:solutionspace} (b) and should be avoided.

The MBB$_{s}$ task adopted five equally separated fixed-position constraints to generate MBB beams supported at intermediate positions, as shown in Figure \ref{fig:solutionspace} (c). The constraint moved from the middle bottom to the right end of the optimization space, labeled as position $[1,2,3,4,5]$. The movement of the constraint was achieved by zeroing the corresponding values in the structure deformation array $\textit{\textbf{U}}$ to be $0$ at the new fixed positions. Unlike the volume constraint, the fixed-position constraint indirectly influenced the loss through the compliance $C$. At the inference stage, smoothly evolving structures were generated, showing NSTO's generation capability with more complex constraints.

The CB$_{v}$ task of generating varying-volume cantilever beams was performed to verify the dual networks' capability in more challenging 3D tasks, as shown in Figure \ref{fig:solutionspace} (d). Within the volume constraint interval of the five subtasks $\delta \in [0.30, 0.35, 0.40, 0.45, 0.50]$, it showed lower compliance than SIMP, which verifies the validity of the solution space generation in 1D latent code scenarios.

\begin{table}[t]\centering
\begin{center}
\caption{Resolution and computation times. $*$ marks the solution space generation tasks.}\label{tbl2}
\resizebox{\linewidth}{!}{
\begin{tabular}{@{}lllllll@{}}
\hline
Object & Resolution & $V/V_0$ & $C/C_0$ & Iter. & FEA(s) & Total(s)\\
\hline
Shelf & $62 \times 15 \times 62$ & $0.26$ & $1.85$ & $71$ & $160.37$ & $160.55$\\
Desk & $39 \times 27 \times 30$ & $0.14$ & $22.71$ & $166$ & $199.92$ & $200.33$\\
Torsion & $20 \times 20 \times 40$ & $0.36$ & $0.23$ & $241$ & $144.54$ & $145.20$\\
Daruma & $18 \times 15 \times 17$ & $0.07$ & $1.00$ & $85$ & $15.32$ & $15.51$\\
Atlas & $32 \times 32 \times 60$ & $0.34$ & $0.44$ & $54$ & $135.71$ & $135.86$\\
Guitar & $240 \times 177$ & $0.76$ & $1.04$ & $311$ & $56.41$ & $57.12$\\
$^*$MBB$_{v}$ & $120 \times 60$ & $[0.20, 0.40]$ & / & $300 \times 5$ & $139.66$ & $146.27$\\
$^*$MBB$_{s}$ & $60 \times 30$ & $0.30$ & / & $300 \times 5$ & $61.45$ & $68.09$\\
$^*$CB$_{v}$ & $40 \times 20 \times 10$ & $[0.30, 0.50]$ & / & $300 \times 5$ & $444.97$ & $495.57$\\
$^*$Drums & $180 \times 1 \times 60$ & $0.30$ & / & $71 \times 8$ & $358.67$ & $364.98$\\
\hline
\\
&\multicolumn{5}{c}{Computation time of NSTO and SIMP in the Daruma task}\\
\hline
\multicolumn{1}{c}{Algorithm} & \multicolumn{1}{c}{NSTO} & \multicolumn{1}{c}{$^1\textrm{SIMP}$} & \multicolumn{2}{c}{$^2\textrm{SIMP}$} & \multicolumn{2}{c}{$^3\textrm{SIMP}$} \\
\multicolumn{1}{c}{Resolution} & \multicolumn{1}{c}{$18 \times 15 \times 17$} & \multicolumn{1}{c}{$18 \times 15 \times 17$} & \multicolumn{2}{c}{$36 \times 30 \times 35$} & \multicolumn{2}{c}{$55 \times 46 \times 53$} \\

\multicolumn{1}{c}{Total(s)} & \multicolumn{1}{c}{$15.51$} & \multicolumn{1}{c}{$16.03$} & \multicolumn{2}{c}{$120.37$} & \multicolumn{2}{c}{$475.56$} \\
\hline
\end{tabular}
}
\end{center}
\end{table}

Finally, the overall computation time of NSTO for various challenging tasks is introduced in Table \ref{tbl2}. Among these tasks, the fastest Daruma doll infill optimization took only $15$s, the complex Atlas sculpture $135.86$s, and the solution space generation of drum shells $364.98$s, which generated $7$ drum shells. For reference, the state-of-the-art research \cite{wu2015system} developed a high-performance GPU solver that optimized \textit{kitten} in $263.33$s, \textit{bunny} in $400.36$s and \textit{Neptune} sculpture in $709.16$s. The Daruma task exemplifies the computation time of NSTO and SIMP. Within 85 iterations, the two methods used up similar times at the same FEA resolution. Nevertheless, the SIMP computation time increased cubically at higher resolutions (i.e., at the $1\times, 2\times, 3\times$ super-resolution) and could have exceeded the VRAM capacity.

In NSTO, assembling the global stiffness matrix $K$ and solving the structure deformation uses up most of the computation time, whereas the other parts have negligible costs. For example, the network's average feedforward and backpropagation time were $0.0645$s and $0.0766$s for a $240 \times 80$ resolution MBB beam optimization. The stiffness matrix assembly time was 0.486s, and the linear elasticity solving time was 0.160s (with AMGX \cite{naumov2015amgx}), taking up around $82.04\%$ of the computation time. Similar time uses of around $80\%$ to $90\%$ were observed at various resolutions.

In conclusion, the convergence, optimization performance, and generation capability are presented and analyzed. Compared with the benchmark SIMP, NSTO demonstrated high optimization quality. The modulator and oscillator networks also demonstrated structure generation capability under direct and indirect varying boundary conditions or constraints.

\subsection{Applications}\label{subsec4.3}
This section demonstrates a wide range of personal fabrication applications with NSTO, including \textit{\textbf{a:}} 2D optimization tasks of printing infills and electric guitar body;  \textit{\textbf{b:}} 3D optimization tasks of the bookshelf, table, torsion, Daruma doll,, and Atlas sculpture; \textit{\textbf{c:}} multi-structure optimization of the jazz drum shells.

\begin{figure}[t]
    \centering
    \includegraphics[width=\linewidth]{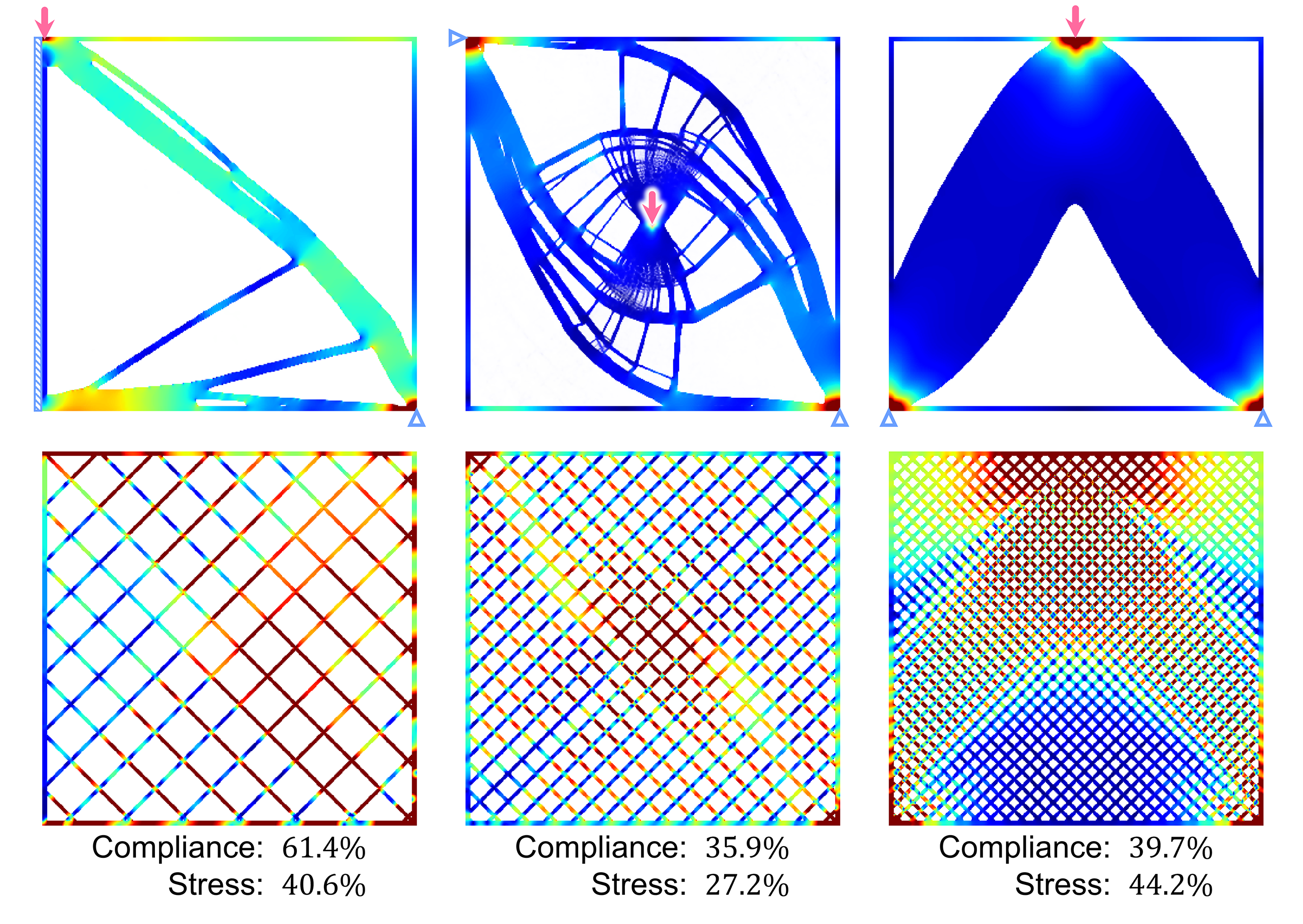}
    \caption{Comparison of the topology-optimized and grid infills under different infill percentages and boundary conditions. The red and blue colors indicate high and low stress, respectively.}
    \label{fig:infill2d}
\end{figure}

\begin{figure}[h]
    \centering
    \includegraphics[width=\linewidth]{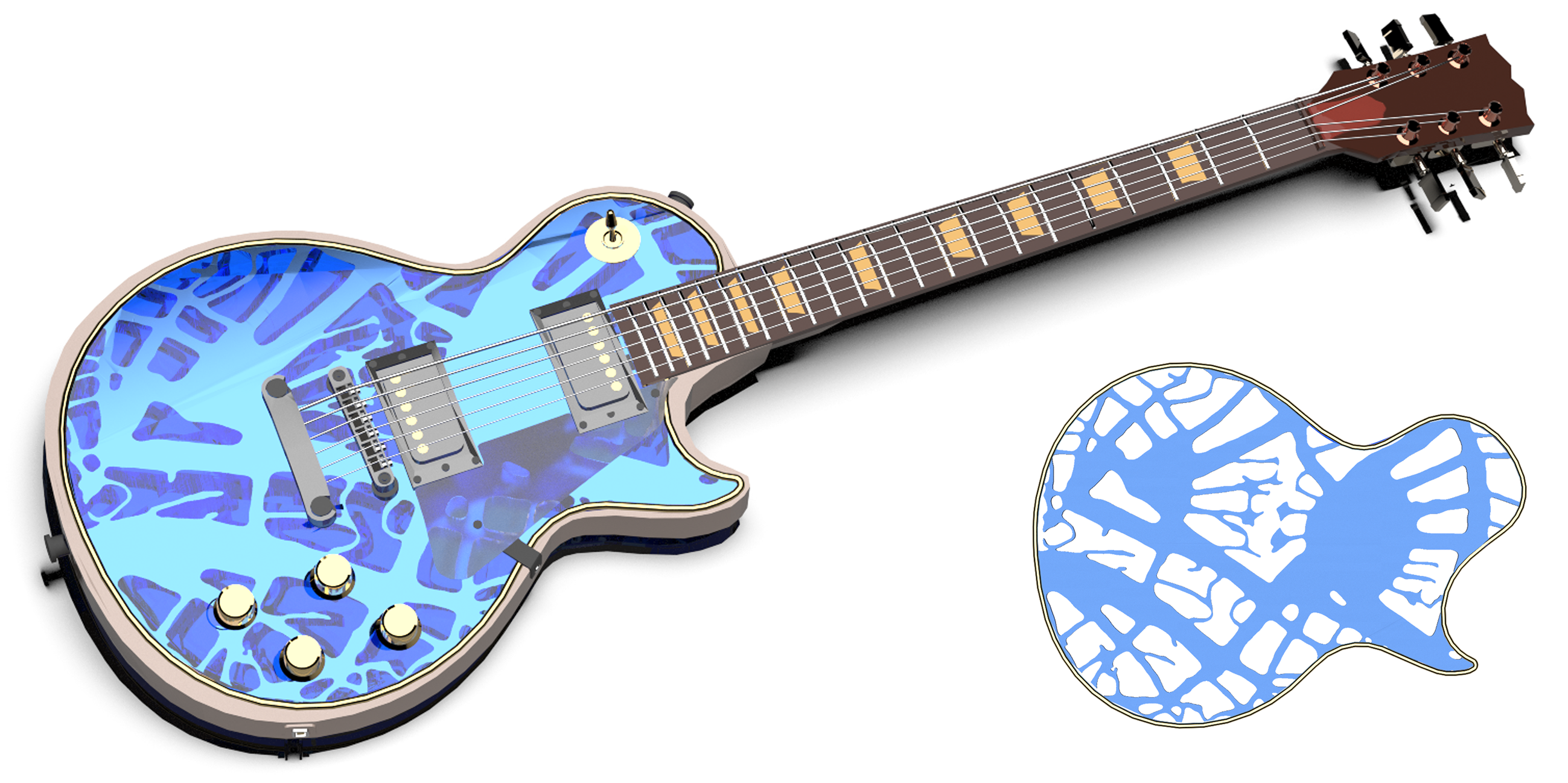}
    \caption{Optimized electric guitar body under planar-distributed loads, featuring spiderweb-like features.}
    \label{fig:guitar}
\end{figure}

\begin{figure}[ht]
    \centering
    \includegraphics[width=0.885\linewidth]{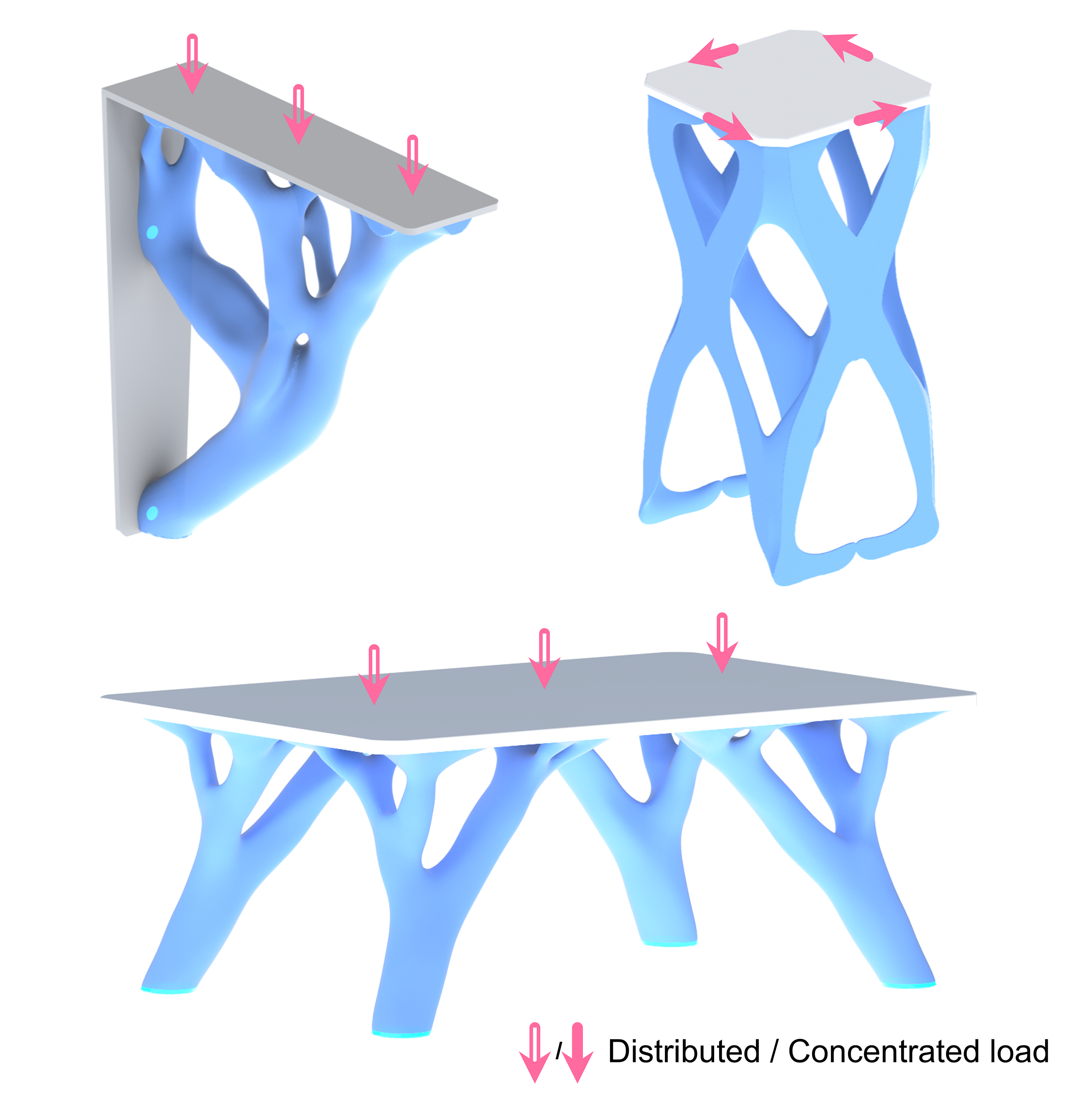}
    \caption{Structure generation results of shelf, desk and torsion structures. Red arrow show the external load's positions and it's type, light blue areas indicate the fixed positions.}
    \label{fig:table-shelf-torsion}
\end{figure}

\textit{3D Printing infill} aims to save materials to ensure structural strength. It is also one of the most practical applications of structural topology optimization in personal fabrication. Here, we compare three pairs of topology optimized infill structures with regular grid infills under different external forces, as shown in Figure \ref{fig:infill2d}. The optimized structures have the same volume fraction (i.e., material consumption) as the grid infill: $22.3\%$, $37.7\%$, and $57.9\%$. Compared with the uniform grid infill, the optimized infill structures showed lower compliance of $61.4\%$, $35.9\%$ and $39.7\%$, and lower mean stress of $40.6\%$, $27.2\%$ and $44.2\%$ respectively, therefore allowing the design of more robust loaded structures.

\textit{Guitar}, flute \cite{umetani2016printone}, metallophone \cite{bharaj2015computational}, and various 3D-printed musical instruments were created by artists and hobbyists because of their unique appearance and acoustic properties. Here, we optimized an electric guitar body under a planar-distributed load, as shown in Figure \ref{fig:guitar}. The 2D optimization results were stretched up and Boolean-intersected with the original body to achieve unique designs. The solution ensured structural stiffness and demonstrated a spider web-like personalized feature, which could be stronger and more attractive than repetitive features.

We reviewed NSTO's performance on non-shape-constraint and complex-shape-constraint tasks with two applications: the bookshelves, table, and torsion tasks, and the Daruma dolls and Atlas sculpture tasks.

\textit{Bookshelf, desk, and torsion} structures were successively optimized, as shown in Figure \ref{fig:table-shelf-torsion}. They were under the representative boundary conditions of the distributed force and torque, and no shape constraints were applied. After the optimization, all the structures were meshed with the marching cube algorithm \cite{lorensen1987marching}.

The shelf was fixed at two points and subjected to a uniformly distributed force. Compared with the concentrated-load cases, the optimization objects under the distributed load tended to grow more high-frequency branches, which indicates the convergence and high-frequency expressiveness of the optimization algorithm. The solution revealed that the thick structures of the shelf were extended to the fixed end, and the thin structures were rooted to the loaded surface, verifying that NSTO can generate sufficient high-frequency structural details in 3D cases.

The desk was subjected to a distributed normal force on the surface and was supported at four corners. For simplicity, mechanical symmetry constraints were applied by fixing the horizontal degrees of freedom of the two symmetry planes (the vertical axis points towards the height direction) to optimize a quarter of the desk. The solution presented a natural dendritic pattern that diffused from the constraint position to the distributed load.

\begin{figure}[t]
    \centering
    \includegraphics[width=0.82\linewidth]{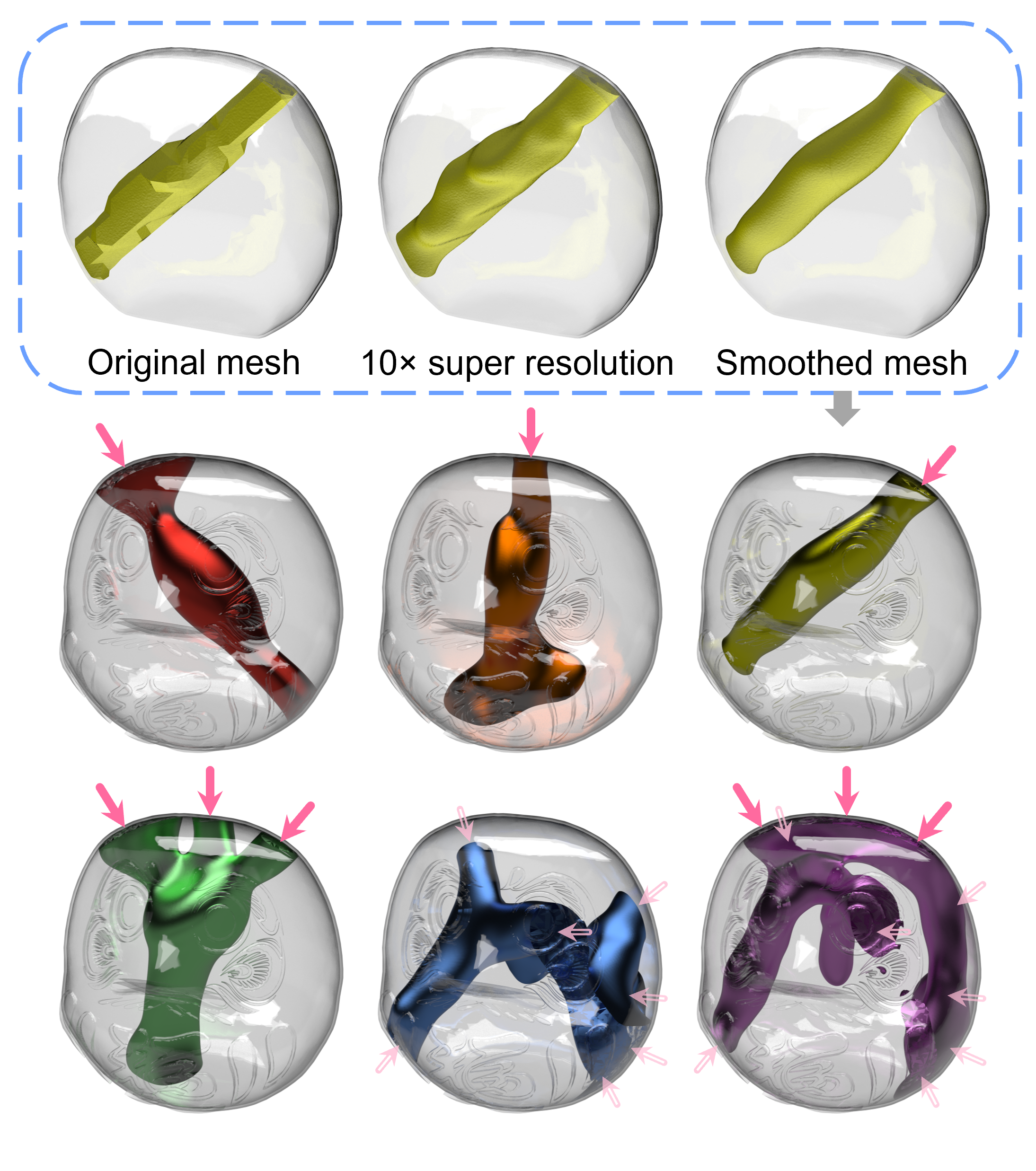}
    \caption{The optimization process of Daruma dolls and the final solutions under various load conditions.}
    \label{fig:daruma}
\end{figure}

The torsion structure was subjected to four circumferential external forces as torques while fixed at the bottom. One-quarter of the structure was symmetrically mapped to the other three parts through shape regulation to form a more aesthetic solution. Thanks to the representation freedom that NSTO provided, the mapping between the structural features can also develop in various ways other than symmetric shape regulation, such as multi-scale structural optimization \cite{zhu2017two} or the personalized topology texture \cite{hu2019texture}. In brief, the above solutions intuitively demonstrated NSTO's capability in non-shape-constraint tasks. 

\begin{figure}[t]
    \centering
    \includegraphics[width=\linewidth]{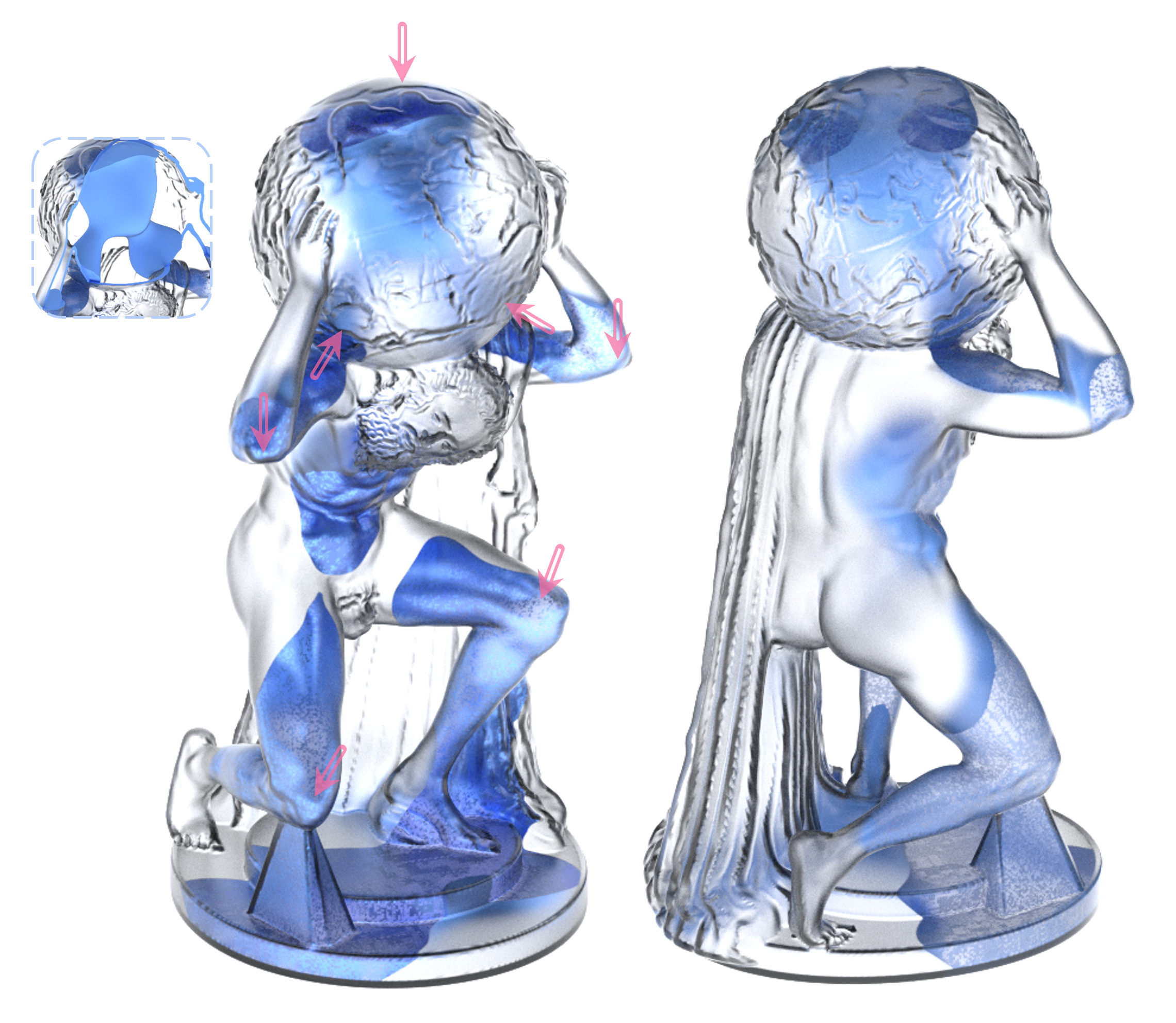}
    \caption{Optimized Atlas sculpture. A section drawing is attached for viewing inside the sculpture.}
    \label{fig:atlas}
\end{figure}

\begin{figure}[ht]
    \centering
    \includegraphics[width=\linewidth]{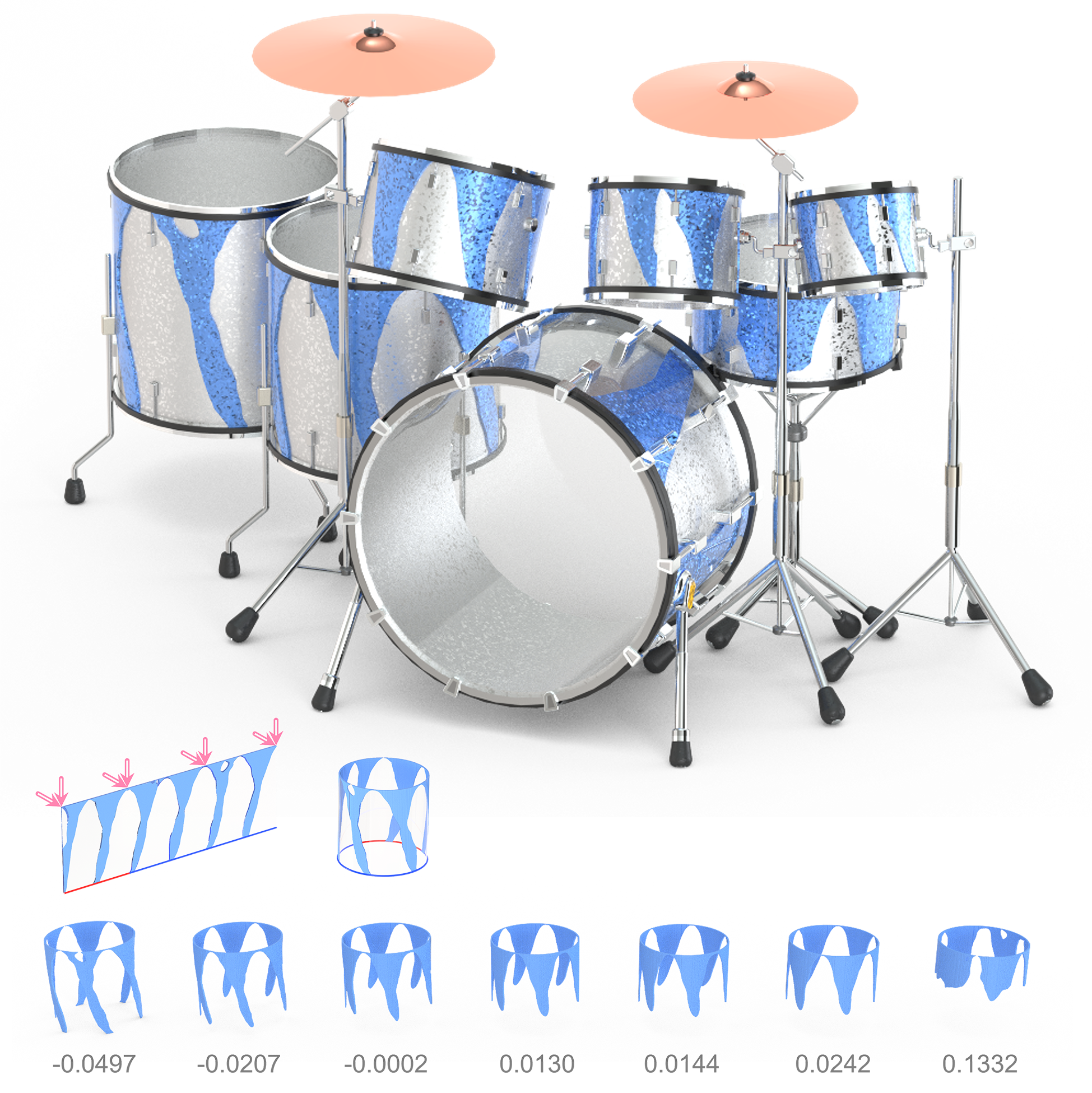}
    \caption{Solution space of drum shells, where diameter height ratio is the varying boundary condition. The top image shows the mapping between annular and planar structures, and the bottom image shows the solutions with latent codes.}
    \label{fig:drums}
\end{figure}

When optimizing structures with shape constraints, the boundary structures are first voxelized into structured grids. Then, the material density values outside the structure are set at $0$. A thin shell is preserved for 3D printing.

\textit{Daruma dolls} were optimized under several groups of external forces, including the concentrated force at the top and the discrete force over the whole body, as shown in Figure \ref{fig:daruma}. NSTO generated an intuitive structure along the direction of the concentrated force, and discrete-force-basis structures are much more complex. The upper part of Figure \ref{fig:daruma} presents the optimized structure with a $10$-times axial super-resolution (i.e., a $1000$-times voxel super-resolution). Then, surface Laplacian smoothing was selectively applied. It should be noted that the Daruma optimization took only around $15$s.

\textit{Atlas sculpture} was selected for the complex-shape-constraint optimization, as shown in Figure \ref{fig:atlas}. Multi-directional external forces were added to its raised sphere and joints. We observed that supporting structures were grown from the contact position to strengthen the local stiffness, proving that NSTO is effective for complex-shape-constraint situations.

Finally, we leveraged the dual networks for the drum shell optimization under a varying volume constraint, as shown in Figure \ref{fig:drums}. \textit{Drum shells} are annular thin-walled structures responsible for carrying the load. While resisting radial tension and axial pressure, they need to be lightweight to retain greater resonance during vibration for producing longer sustain, which is favored by drummers. To this end, the drum shells were optimized under volume constraints, radial tension, and axial pressure, where the diameter height ratio was set as the dynamic boundary condition to find the optimal design scheme for the drums of different sizes. The solutions will be used as the lightweight outer layer of the drum shell to improve the structural stiffness without affecting the inner air vibration.

\begin{figure*}[t]
    \centering
    \includegraphics[width=\linewidth]{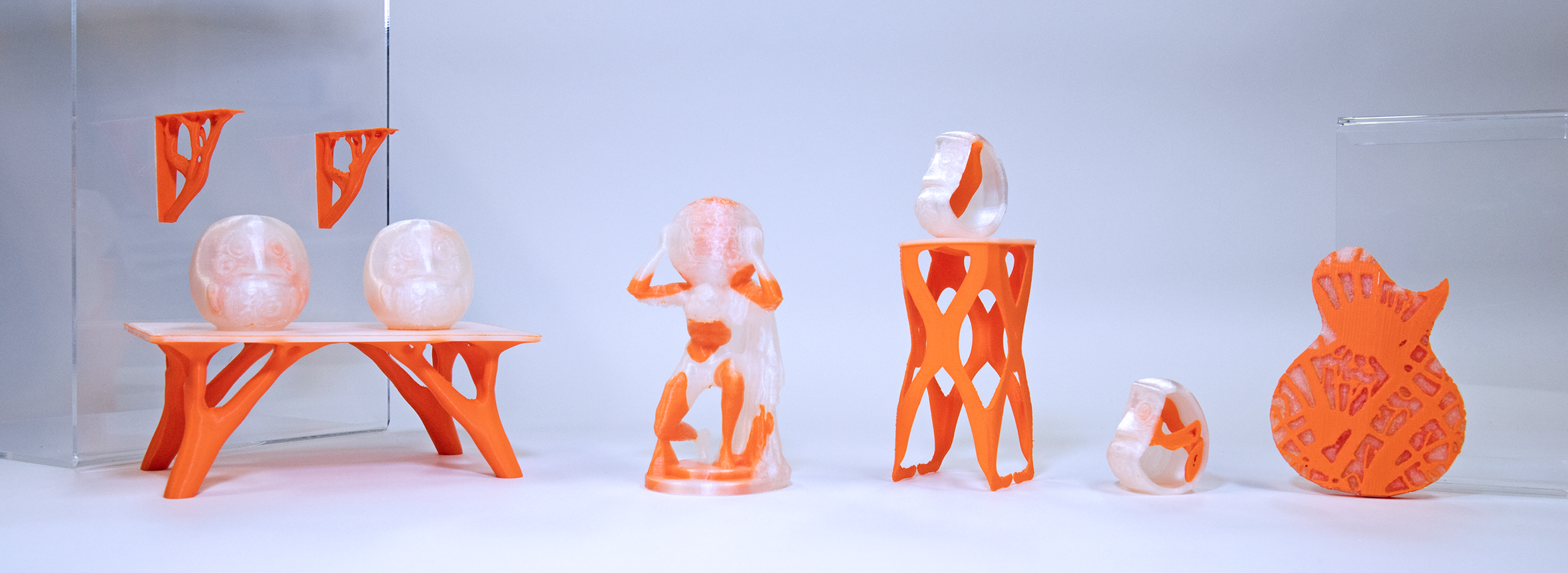}
    \caption{3D-printed topology-optimized structures using the fused deposition modeling 3D printer (Ultimaker 3).}
    \label{fig:photo}
\end{figure*}

The diameter height ratio of the drum shell ranged from $5.5/14$ for the snare drum to $16/16$ for the floor tom. Therefore, we used 1D latent code and optimized the solutions through $8$ subtasks with diameter height ratios uniformly distributed between $[0.3, 0.7]$. To make more efficient use of the optimization space, the shell was mapped into a $180 \times 1 \times 60$ flat plate. The bottom constraints were evenly moved upward to modify the diameter height ratio. At the inference stage, users may slide along the 1D latent space at the inference stage and freely select the desired solution according to each drum's diameter height ratio.

The topology-optimized structures were printed with the fused deposition modeling printer to verify the printability, as Figure \ref{fig:photo}.

\subsection{Limitations and future works}\label{subsec4.4}

The selection and refinement of the structure discretization influenced the super-resolved results of the optimized structure. Specifically, when optimizing structures at a single resolution, the network may overfit current solutions after extensive training, thus causing resolution-dependant artifacts such as zigzag structure boundaries due to the rectangular finite elements. A potential solution is to compute the structural performance with random numbers, such as through Monte Carlo integration, although the solving efficiency provided by the grid elements will be lost \cite{liu2018narrow}.

\setlength{\columnsep}{3pt}
\begin{wrapfigure}{r}{5.8cm}
\label{wrap-fig:3}
\vspace{-4mm}
\hspace{-0.5mm}
\includegraphics[width=\linewidth]{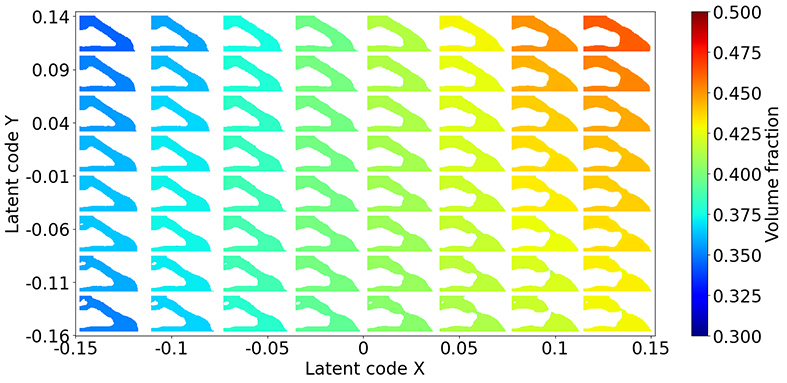}
\vspace{-8mm}
\end{wrapfigure}

Our future work is to explore the strategy of solution spaces optimized under multiple boundary conditions or constraints. Currently, we implemented a 2D solution space under varying volume constraints and supporting positions. The solution space demonstrates a radiative pattern of constraints, calling for more efficient trajectory planning strategies and methods for imposing varying boundary conditions or constraints in high-dimensional situations.

The relationship between network frequency features and the structural performance at various resolutions can be the topic of another future work. In NSTO, the network frequency hyperparameter $\omega$ has a positive relationship with the structure details, resulting in compliance change. Viewing and optimizing the structural performance from the Fourier space is a potential angle.

\section{Conclusions}\label{sec5}

Previously, most structural topology optimization frameworks were for pure numerical research or industrial applications, leaving a gap between the requirements of ordinary users and algorithm developments. Therefore, we proposed the NSTO framework, which provides optimal structures conducive to manufacturing at a lower computational cost and enables novel structure generation.

NSTO significantly improves the optimization efficiency from the resolution aspect to ensure optimal performance. It also expands the solution space purely under the physical constraints of multiple subtasks through self-supervision. The optimization results under complex boundary conditions are mapped to a latent space, so users may infer novel solutions without repeating the topology optimization process. With the injection of creativity, users are expected to optimize sets of functional structures for decoration or interaction. We supported the above claims through a comprehensive study of the algorithm performance, comparison with the benchmark method, and examination of a wide range of tasks with complex geometric constraints and boundary conditions.

\section*{Acknowledgements}
We are grateful to the anonymous reviewers for their comments. This work was supported by JSPS KAKENHI Grant Number JP19K20321 and JP20H05958, JST PRESTO Grant Number JPMJPR19J2, and JST ACT-X Grant Number JPMJAX20AK.

\printbibliography               
\newpage

\end{document}